\documentclass[aps,prb,twocolumn,superscriptaddress,longbibliography,notitlepage]{revtex4-1}

\usepackage{graphicx}
\usepackage{amsmath}
\usepackage{amstext}
\usepackage{amssymb}
\usepackage[colorlinks,citecolor=blue]{hyperref}
\usepackage{graphicx}
\usepackage{amsmath}
\usepackage{amstext}
\usepackage{amssymb}
\usepackage{amsfonts}
\usepackage{longtable,booktabs}
\usepackage{hyperref}\usepackage{url}
\usepackage{subfigure}%
\usepackage{dsfont}

\usepackage{amsbsy}
\usepackage{dcolumn}
\usepackage{amsthm}
\usepackage{bm}
\usepackage{esint}
\usepackage{multirow}
\usepackage{hyperref}
\hypersetup{
    colorlinks=true,
    linkcolor=blue,
    filecolor=magenta,
    urlcolor=blue,
}
\usepackage{cleveref}

\usepackage{mathrsfs}
\usepackage{amsfonts}
\usepackage{amsbsy}
\usepackage{dcolumn}
\usepackage{bm}
\usepackage{multirow}
\usepackage{color}
\def\Z{\mathbb{Z}}

\def\ep{\epsilon^{\mu\nu\lambda\rho} }
  \usepackage{extarrows}
 
\usepackage{datetime}

     \newtheorem{thm}{Theorem}

       \newtheorem{dfn}{Definition}
       \newtheorem{crt}{Criterion}
      \theoremstyle{remark}

\begin{document}

 \title{{\fontfamily{ptm}\fontseries{b}\selectfont Composite Particle Theory of Three-dimensional Gapped Fermionic Phases:   Fractional Topological Insulators and Charge-Loop Excitation Symmetry }}

\author{Peng Ye}
   \affiliation{Department of Physics and Institute for Condensed Matter Theory,
University of Illinois at Urbana-Champaign, IL 61801, USA}
 
 \author{Taylor L. Hughes}
   \affiliation{Department of Physics and Institute for Condensed Matter Theory,
University of Illinois at Urbana-Champaign, IL 61801, USA}

 \author{Joseph Maciejko}
    \affiliation{Department of Physics, University of Alberta, Edmonton, Alberta T6G 2E1, Canada}
   \affiliation{Theoretical Physics Institute, University of Alberta, Edmonton, Alberta T6G 2E1, Canada}
   \affiliation{Canadian Institute for Advanced Research, Toronto, Ontario M5G 1Z8, Canada}
\author{Eduardo Fradkin}
   \affiliation{Department of Physics and Institute for Condensed Matter Theory,
University of Illinois at Urbana-Champaign, IL 61801, USA}

\begin{abstract}  
{\fontfamily{ptm}\fontseries{c}\selectfont  
{Topological phases of matter are usually  realized in deconfined phases of gauge theories. In this context, {confined} phases with strongly fluctuating gauge fields  seem to be irrelevant to the physics of topological phases.} For example, the low-energy theory of the two-dimensional (2D) toric code model {(i.e. the deconfined phase of $\mathbb{Z}_2$ gauge theory)} is a $U(1)\times U(1)$ Chern-Simons theory in which gauge charges (i.e., $e$ and $m$ particles) are deconfined and the gauge fields are gapped,   while the confined phase is topologically trivial. In this paper, we point out a new route to {constructing exotic 3D gapped  fermionic} phases {in a confining phase of a gauge theory}. {Starting from a parton construction with strongly fluctuating compact $U(1)\times U(1)$ gauge fields, we construct gapped phases of interacting fermions} by condensing two linearly independent bosonic composite particles {consisting} of partons and  $U(1)\times U(1)$ magnetic monopoles. {This} can be regarded as a 3D generalization of the 2D Bais-Slingerland condensation mechanism. Charge fractionalization {results} from a Debye-H\"uckel-like screening cloud formed by the condensed composite particles. {Within our general framework, we explore two aspects of symmetry-enriched} 3D Abelian topological phases. First, {we construct a new fermionic state of matter with time-reversal symmetry and $\Theta\neq \pi$, the fractional topological insulator}. Second, we generalize the notion of \emph{anyonic symmetry} of 2D Abelian topological phases to the \emph{charge-loop excitation symmetry} ($\mathsf{Charles}$) {of} 3D Abelian topological phases. {We show that line twist defects, which realize $\mathsf{Charles}$ transformations, exhibit non-Abelian fusion properties.}}
 \end{abstract}


  \maketitle

\section{Introduction}
   \subsection{Background and overview: Parton construction and gauge confinement}
 In models of non-interacting fermions, several topological phases of matter have been found, such as integer  quantum Hall states (IQH), Chern insulators, and topological insulators (TI)\cite{iqh,haldane88,TI1,TI2,TI3,TI4,TI5,TI6,TIexp}.    Owing to the non-interacting nature of the problem, tremendous progress has been made in both theory and experiment. In the presence of  weak interactions these phases can also be analytically understood.  If interactions are strong enough such that a perturbative analysis is no longer meaningful,  one usually faces a problem in  strongly-correlated electron physics. While exact solutions are possible in a few specific models, one often constructs approximate effective descriptions of such systems.
One such approach is the parton construction approach, also known as the projective construction, or  slave-particle approach. It has been  widely applied in   studies of strongly correlated electron systems such as high-temperature superconductors and fractional quantum Hall states (FQH).\cite{BZA8773,BA8880,AM8874,KL8842,SHF8868,AZH8845,DFM8826,WWZcsp,Wsrvb,LN9221,MF9400,WLsu2,WenRMP,weng,ye10,ye11a,ye_zhang,ye_ma,ye_wang,jain_parton,wen_parton,Wen99,Barkeshli2010} Recently, it has also been applied \cite{LL1263,YW12,YW13a,Wangsenthil2015,Ye14b}   to bosonic symmetry-protected topological phases (SPT) as well.\cite{1DSPT,Chenlong,Chen_science,Chen10}

Generally speaking, in the parton construction we start with a lattice action $S$ that describes a strongly-correlated electronic system. In this paper, the electron operator $c$ is meant to represent a  generic Grassmann variable that is the only dynamical variable in $S$. We further write the electron operator $c$ in terms of several parton operators $f^i$. The Hilbert space for $S$ is equivalently replaced by a projected Hilbert space formed by partons and gauge fields. In practice, there are many different kinds of parton constructions. We focus on one of them, where  all partons are fermionic such that an odd number of partons is required to form an electron. Mathematically, the electron operator is formally fractionalized as  $c=f^1f^2\cdots f^{2n+1}$.  The electron operator 
 $c$  is  a singlet of the $SU(2n+1)$ gauge group. The largest totally commuting subgroup, or maximal torus, is given by the compact Abelian $(U(1))^{2n}$ gauge group which acts with gauge transformations: $f^1\rightarrow f^1e^{i\theta_1},\cdots, f^i\rightarrow f^ie^{i\theta_{i}-i\theta_{i-1}},\cdots, f^{2n+1}\rightarrow f^{2n+1}e^{-i\theta_{2n}}$, where $\{\theta_i\}$ ($i=1,2,\cdots,2n$) are arbitrary  functions of the lattice sites and a continuous time variable. As such,  by applying the 't Hooft gauge projection,\cite{Gauge_Confinement_tHooft3} the lattice action deep in the confined phase is reformulated to describe a system of interacting  partons and $2n$ dynamical compact abelian gauge fields $\{a^{(i)}_\mu\}$.

  It should be noted that  the gauge-field coupling constants $g_i$  at the lattice scale should be treated as being very strong since  the usual lattice kinetic  terms (with coefficient $1/g^2_i$) for compact gauge fields are not present in $S$. 
From here one can usually proceed further by assuming a mean-field theory of partons where the effects of gauge fluctuations are assumed negligible. As such, a very important feature---the compactness of gauge fields---is totally ignored. A standard perturbative analysis can be applied in order to quantitatively recover the   effect of gauge fluctuations at leading order. In some cases, this assumption is legitimate. A typical example is a 2D system where fermionic partons occupy energy bands with non-zero Chern number at the mean-field level.\cite{YW12} In this case, a Chern-Simons term is generated and a topological mass gap\cite{csgap} for the gauge fields is produced, which suppresses  instanton tunneling. However, there is no reason to rule out the possibility that $g$ at low energies flows to  strong coupling, such that  the compactness of the gauge fields plays  a   fundamental role in reshaping the nature of the emergent ground states. In such cases,    mean-field theories of partons fail  to describe the physical states formed by electrons, even at  a qualitative level.

Despite  the strong coupling nature of the problem,  the leading order effect  of gauge fluctuations can still be perturbatively treated by considering the Bose-Einstein condensation of composite particles.  In the regime of strong gauge coupling, condensed composites contain magnetic monopoles of the internal compact gauge fields (and possible electric gauge charge as well). Historically, this line of thinking was developed in the context of strongly coupled gauge theories. For example, condensed monopole phases are relevant for studies of  (3+1)D compact quantum electrodynamics, the Georgi-Glashow models, and supersymmetric Yang-Mills theory.\cite{mandelstam,mt78,Gauge_Confinement_Polyakov_1,Gauge_Confinement_Polyakov_2,Gauge_Confinement_Polyakov_3,Gauge_Confinement_tHooft1,Gauge_Confinement_tHooft2,Gauge_Confinement_susskind,Gauge_Confinement_Fradkin,Gauge_Confinement_tHooft3,Gauge_Confinement_proceedings,susy}  For an Abelian gauge theory with a compact $U(1)$ gauge group, the monopole creation operator has been constructed explicitly and gains a nonzero vacuum expectation value as shown in Ref.~\onlinecite{Gauge_Confinement_Fradkin}.   Recently, it was further suggested that  the behavior of the non-trivial line operators  may be used to make the proper distinction between confinement phases of strongly coupled gauge theories.\cite{seiberg_order_2013} Note that since electric (i.e., gauge) charge excitations are linearly confined during this condensation process, the monopole condensation phase is also called  the confinement phase.

  Unfortunately, the usual monopole condensation scenario, \cite{mandelstam,mt78,Gauge_Confinement_Polyakov_1,Gauge_Confinement_Polyakov_2,Gauge_Confinement_Polyakov_3,Gauge_Confinement_tHooft1,Gauge_Confinement_tHooft2,Gauge_Confinement_susskind,Gauge_Confinement_Fradkin,Gauge_Confinement_tHooft3,Gauge_Confinement_proceedings,susy} when applied to the 3D parton construction, will simply confine all partons back into  electrons, {resulting in} a non-fractionalized trivial insulator. In this sense, the parton construction with gauge confinement {driven} by the usual monopole condensation {does not seem to be a good pathway to reach topological states}. To save the parton construction approach, we should look for new scenarios of gauge confinement. More precisely,  can we have a new kind of condensation that confines partons while still {leading to} a fractionalized insulator? How can we imagine the existence of fractionalized excitations  when  partons are confined? If these questions can be solved,  a new systematic treatment of 3D fermionic fractionalized phases will be established.  {This is the main goal of this paper, and that this is possible can be gleaned from the success  of  the 2D Bais-Slingerland condensation mechanism}.\cite{bais2009} Indeed, we show that  there are  new pathways to fractionalization in 3D, now in the {\em confining regime} of the gauge theory, provided that confinement occurs as the result of condensation of a class of  composites made of fermionic partons and monopoles (from different sectors of the gauge group). We will see that this new form of {\em oblique confinement}\cite{Gauge_Confinement_tHooft3} leads to unexpected phases of matter, particularly states with fractionalized $\Theta$ angles and yet compatible with time-reversal invariance. {While oblique confinement as a pathway to topological phases has been considered before in the context of bosonic phases of matter,\cite{vonkeyserlingk2015} here we focus on topological phases of interacting fermions.} In addition,  we  also  study several applications. For example, how can we impose  symmetry in such a parton construction with gauge confinement? The latter leads to the notion of symmetry-enriched topological phases (SET) in the parton construction approach.

\subsection{ {Summary of main results}}

  {\emph{\textbf{(1)} Composite particle theory of fermionic phases}}.  In this article, we will consider the condensation of   ``composites''   that  not only carry magnetic charges but also contain fermionic partons that are charged  under different internal gauge fields [i.e., $U(1)\times U(1)$ strongly fluctuating gauge fields in our concrete example $c=f^1f^2f^3$] and under the  external electromagnetic (EM) field $A_\mu$.  One caveat is that, 
 despite the mixture of partons and magnetic monopoles, those condensed composites are \emph{not} dyons. More concretely, they carry either electric charge or magnetic charge in a given gauge group, not both.  This fact allows us to make reliable statements and calculations from a \emph{local}  theory. All excitations can be organized as a set of   \emph{charge-loop composites}, and, as a whole, form a \emph{charge-loop-lattice} in which each lattice site corresponds to a deconfined excitation.  Especially, partons are confined as usual but some composites constituted by partons and magnetic monopoles of $U(1)\times U(1)$ gauge fields may be deconfined and carry fractionalized EM electric charge. Many universal physical properties can be easily determined from the charge-loop-lattice, such as  the  braiding statistics between point-like excitations and loop excitations, the self-statistics of point-like excitations, the EM charge of the excitation, and the bulk axion  $\Theta$ angle.\cite{witten1,Qi2008}  We will refer to this approach to constructing 3D fermionic gapped phases  as a \emph{composite particle theory}.

In this theory,  charge fractionalization  is achieved via a Debye-H{\"u}ckel-type charge-screening cloud formed by the composite condensates. 
This   is analogous to the  charge-screening phenomenon in the composite fermion theory of the FQH effect.\cite{composite_Jain_1,composite_Jain_2,composite_Fradkin,composite_HLR,composite_sm1,cmposite_H_2,composite_sm2}  In fact, we prove that the Debye-H{\"u}ckel-type screening is the unique source of   charge fractionalization.   In principle, all physical quantities of the resulting phases  can be expressed as functions of a set of parameters that characterize composite particle theory. This line of thinking is also analogous to the  composite fermion theory of FQH states where the filling fraction is unified in a sequence of discrete numbers, each of which corresponds to a specific ansatz  in the composite fermion construction. From this perspective, the composite particle theory may be regarded as an attempt to find a 3D analog of the composite fermion theory of FQH states{, with the caveat that we are considering confined phases while in the composite fermion theory, all gauge fields are deconfined. }

 {\emph{\textbf{(2)}   Fractional topological insulators}}. 
Based on the composite particle theory,  we will study two symmetry-enriched properties. The first property is the bulk axion  angle $\Theta$ in the presence of time-reversal symmetry.\cite{witten1,Qi2008} When $\Theta$ is non-vanishing an externally inserted EM monopole  with integral magnetic charge $M$ will induce an electric polarization charge   $\frac{\Theta}{2\pi}M$,     a phenomenon known as the Witten effect.\cite{witten1,Qi2008,franz} For free-fermion topological insulators, the $\Theta$ angle is $\pi \text{ mod } 2\pi$ with $M\in\Z$.\cite{Qi2008}  However, it was theoretically proposed   that   $\Theta$ could be   different from $\pi$ if strong interactions and correlations are taken into account,\cite{maciejkoFTI,maciejko_model,maciejko2015,swingle2011,swingle_fti_2012} leading to the  notion of 3D fractional topological insulators (FTI) with deconfined gauge fields. The periodicity of $\Theta$ should also be properly modified    so as to preserve time-reversal symmetry.

 {In Ref.~\onlinecite{maciejkoFTI}, {FTI}s were obtained via parton constructions  where the internal gauge fields are in the Coulomb phase (photons are gapless). Therefore, a gapless channel, despite of its electric neutrality, can in principle adiabatically connect the FTIs to a fractionalized state with vanishing axion angle. In Ref.~\onlinecite{swingle2011}, bosonic FTIs were obtained via parton constructions where the mean-field Hamiltonian of partons explicitly breaks SU(2) gauge group down to $\Z_2$ discrete gauge group. As a result, the unbroken discrete gauge group leads to bulk topological order and deconfined fractionalized excitations. The gauge fluctuations in both Ref.~\onlinecite{maciejkoFTI} and Ref.~\onlinecite{swingle2011} are perturbatively weak.  In the present work, we  explore the possibility of realizing FTIs via the  condensation of composites (introduced above) when gauge fluctuations are sufficiently strong and gauge confinement occurs. }

 { In Ref.~\onlinecite{maciejkoFTI} and Ref.~\onlinecite{swingle2011},  each parton is assumed to carry a fractional EM electric charge such that a fractionalized $\Theta$ angle should be expected (by simply noting that the coefficient of $F\wedge F$ has unit of $e^2$). This is not the case in our work. We show that even if partons   carry integral EM electric charge (i.e., both $f^1$ and $f^2$ carry $+1$ electric charge and $f^3$ carries $-1$  eletric charge, see Sec.~\ref{sec:bec} for more details), a  fractionalized $\Theta$ and gapped bulk can  also be achieved as long as a proper composite condensation is considered and partons occupy non-trivial topological insulator bands. This feature is unique in the parton construction with gauge confinement.}

 { In the {FTI} state constructed in this work (Sec.~\ref{sec_FTI}, \ref{sec_FTI_plus}), we show that the EM electric charge of deconfined excitations is fractionalized at $1/3$. This is consistent to the claim by Swingle, \textit{et. al.}\cite{swingle2011} that an FTI necessarily has a fractionalized bulk. Indeed, the fractionalization nature of the FTI state in the present work can be traced back to the presence of $\Z_2\times \Z_6$ topological order (see Sec.~\ref{sec_FTI_plus} for more details). The latter arises from the deconfined discrete   subgroup  of  the confined $SU(3)$ gauge group.}

 {\emph{\textbf{(3)} Charge-loop excitation symmetry and extrinsic twist defects}}.  Noting that the set of all excitations forms a charge-loop-lattice,  the  second symmetry-enriched  property is the  concept of  ``charge-loop excitation symmetry'', abbreviated as $\mathsf{Charles}$ (see Definition \ref{dfn_charlos}).  $\mathsf{Charles}$ can be viewed as a hidden symmetry of (3+1)D topological quantum field theories. Meanwhile, $\mathsf{Charles}$ has a geometric interpretation as a point-group symmetry of the charge-loop-lattice that preserves  physical properties of the excitations.  The study of  $\mathsf{Charles}$ is  motivated by the theory of  anyonic symmetry \cite{Teo2015,Teo2014,ran,barkeshli_wen,Teo2013,Barkeshli2014,You2012,genon_1,genon_2,genon_3,genon_4,Bombin2010} and its relation to extrinsic twist defects of 2D Abelian topological phases.   We expect that  3D Abelian topological phases where  charge-loop composite excitations are allowed may host even more exotic physics if extrinsic defects are considered.

      Physically, extrinsic defects  (which may come in the form of vortices or disclinations, for example) are semi-classical objects that are externally imposed into a 2D  topological phase. An extrinsic \emph{twist} defect is one which may be associated with an element of an anyonic symmetry group that acts to permute the set of anyons.  The inclusion of such defects enriches the tensor category theory of the Abelian parent  topological phase.  Indeed, this line of thinking   has attracted  a lot of attention since extrinsic twist defects  can  bind non-Abelian objects even though   all of   excitations of the parent  topological phase without defects are Abelian.\cite{genon_1,genon_2,Bombin2010,genon_3,genon_4,You2012}  A typical example is found in some lattice systems exhibiting $\mathbb{Z}_N$ topological order that contain, for example, the   $\Z_N$ charge and flux anyons $e$ and $m$.\cite{Bombin2010,You2012} In these cases  the anyonic symmetry is intertwined with a lattice translation symmetry such that a dislocation defect acts to exchange the $e$ particle-type with the $m$ particle-type when they orbit around the defect. This implies that the defect harbors a rich internal (non-Abelian) structure so that it can convert between the anyon types.  In the present work we propose   $\mathsf{Charles}$ as  a 3D version of  anyonic symmetry in 2D. In analogy to 2D, each extrinsic defect in 3D is also associated with a $\mathsf{Charles}$ group element. We also study    defect species and some defect fusion properties  (see Fig.~\ref{figure_defect_composite}).

 The remainder of the paper is organized as follows. Sec.~\ref{sec:gauge_strc} is devoted to a general discussion of the parton construction and composite condensation. In Sec.~\ref{sec:theta}, {FTI}s are constructed from a  composite condensation phase where all of the partons occupy topological insulator bands. A concrete example with time-reversal symmetry and fractional $\Theta=\frac{\pi}{9}\text{  mod  }\frac{2}{9}\pi$  is shown (see Fig.~\ref{figure_theta}, Sec.~\ref{sec:theta}). As a comparison, we also show a parton construction in the Coulomb (gapless photon) phase using a perturbative approach, which leads to $\Theta=\pi\text{  mod  }2\pi$ and two gapless neutral modes in the bulk. In Sec.~\ref{sec:charlos_defect_symmetry}, the charge-loop excitation symmetry ($\mathsf{Charles}$) of the charge-loop excitations, and its  relation to 3D extrinsic defects, is studied.  Sec.~\ref{sec:conclusion_direction} is devoted to  the conclusion and future directions. Many key notations, mathematical formulae, and terminologies are introduced in Sec.~\ref{sec:gauge_strc}, which provides the preliminaries for the subsequent parts. 
  Several technical details can be found in the  Appendices.  In  Appendix \ref{appendix:notation}, several notations and abbreviations are collected.


\section{Composite particle theory in three dimensions: A general discussion}\label{sec:gauge_strc}
    \subsection{Compact $U(1)\times U(1)\times U(1)$ gauge symmetry of composites}\label{sec:bec}

    \begin{figure}[t]
\centering
\includegraphics[width=8.5cm]{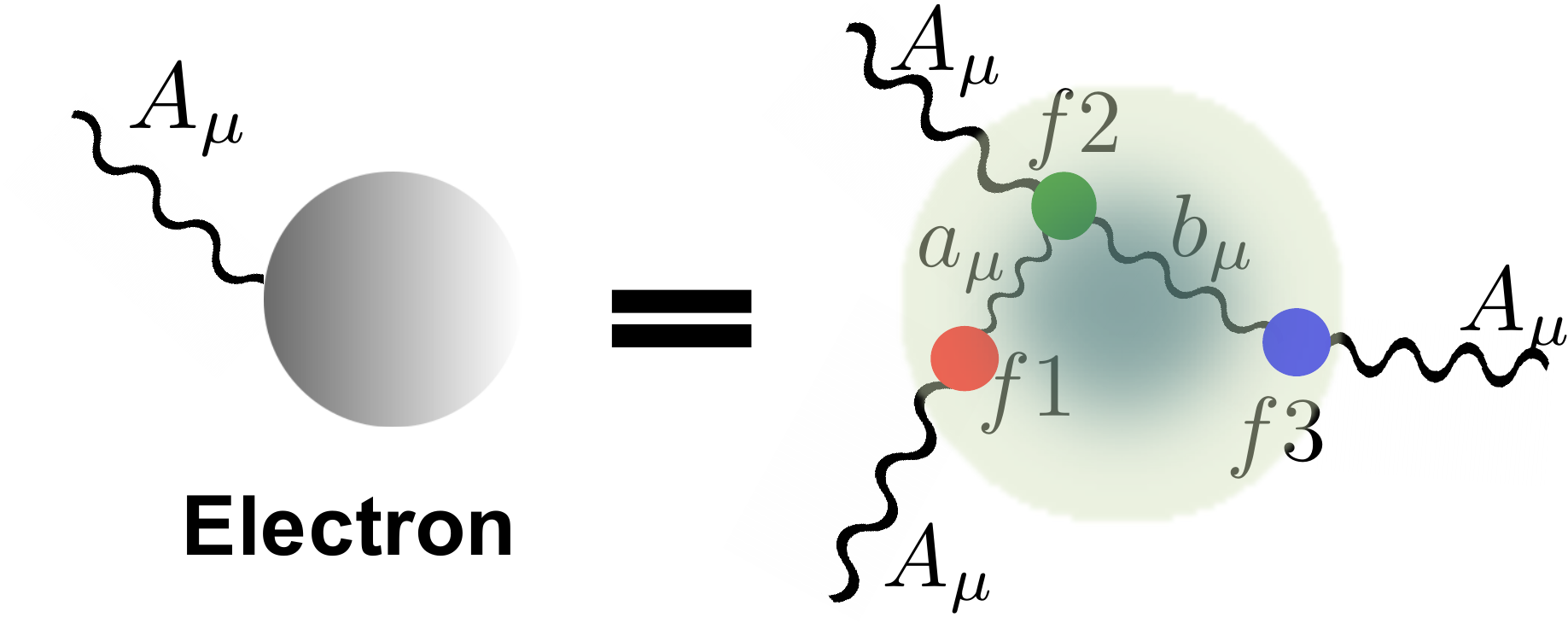}
\caption{(Color online) Parton construction of electron operators in the present work. The wavy lines denote interactions mediated by gauge bosons. $A_\mu$ is the external non-dynamical EM (electromagnetic) field, serving as a probe  of the electromagnetic response of the system. $a_\mu$ and $b_\mu$ are two dynamical, compact $U(1)$ gauge fields, belonging to the $U(1)_a$ and $U(1)_b$ gauge groups respectively. The partons $f^1$ and $f^2$ carry $1$ and $-1$ gauge charges of the  $U(1)_a$ gauge group respectively. The partons $f^3$ and $f^2$ carry $1$ and $-1$ gauge charges of the $U(1)_b$ gauge group respectively. The EM electric charges carried by the partons $f^1$, $f^2$, $f^3$ are $e$, $e$, $-e$, respectively.}
\label{figure_newelectron}
\end{figure}
  
 In the simplest fermionic parton construction,  the   electron operator is decomposed into three fermionic partons: $c=f^1f^2f^3$ (Fig.~\ref{figure_newelectron}), where both $f^1$ and $f^2$ carry unit EM charge $e$ while $f^3$ carries $-e$. As a result, the electron carries $e$.   
In a 3D mean-field theory where the partons are deconfined, we consider that  all fermionic partons have a gapped spectrum and form either a trivial band insulator ($\theta=0$) or a strong topological insulator ($\theta=\pi$) simultaneously, where  $\theta$  denotes the axion angle of the partons \cite{witten1,Qi2008}. To avoid confusion, we will use the capital letter  $\Theta$ to denote the axion angle of the electron, which will be calculated in detail in Sec.~\ref{sec:theta}.   The   internal gauge group is $SU(3)$ whose maximal torus (maximal commuting subgroup)  $U(1)_a\times U(1)_b$ is sufficient to capture the confinement phase properties due to the 't Hooft gauge projection \cite{Gauge_Confinement_tHooft3}.  Here, both $U(1)$ factors are compact gauge groups that support magnetic monopoles \cite{Gauge_Confinement_tHooft3}.  $U(1)_a$ corresponds to the gauge field $a_\mu$ that glues $f^1$ and $f^2$ together, while $U(1)_b$ corresponds to the gauge field $b_\mu$ that glues $f^2$ and $f^3$ together (Fig.~\ref{figure_newelectron}).   Adding  the EM gauge group with gauge field $A_\mu$, the total gauge group is given by   $U(1)_a\times U(1)_b\times U(1)_{\rm EM}$. It should be noted that $A_\mu$ is a non-dynamical (i.e., background) gauge field, which is useful for  diagnosing  the EM linear response properties of the resulting phases. For the same reason, we consider  monopole configurations of $A_\mu$ (with magnetic charge $M$) as externally imposed background configurations. 

 Alternatively, one may also define the following three gauge fields: 
\begin{align}
 A^{f1}_\mu=a_\mu+A_\mu ,  A^{f2}_\mu=-a_\mu-b_\mu+A_\mu , A^{f3}_\mu=b_\mu-A_\mu\,,\nonumber
 \end{align} 
 where $A^{fi}$ is the gauge field that only couples to $f^i$ ($i=1,2,3$). The relation between the two sets of gauge fields can be expressed in matrix form:
\begin{align}
\left(\begin{matrix}
A_\mu\\
a_\mu\\
b_\mu\end{matrix}\right)=\left(\begin{matrix}
1&1&1\\
0&-1&-1\\
1&1&2
\end{matrix}\right)
\left(\begin{matrix}
A_\mu^{f1}\\
A_\mu^{f2}\\
A_\mu^{f3}
\end{matrix}\right)\,,\label{transtion_matrix_gauge}
\end{align}
where the matrix is integer-valued and invertible, i.e., belongs to the $\mathbb{GL}(3,\Z)$ group. 
 We now turn to the description of generic composite particles, which are labeled by a set of electric charges and magnetic charges. We use $N_{a,b}$ to denote the electric charge of the $U(1)_{a,b}$ gauge group and $N_m^{a,b}$ to denote the magnetic charge of that same gauge group. We use $N_A$ and $M$ to denote the bare electric and magnetic charges in the EM gauge group, and $N^{fi}$ and $N^{fi}_m$ to denote the electric and magnetic charges of the $U(1)_{fi}$ gauge groups.
 
 Due to Eq.~(\ref{transtion_matrix_gauge}), the magnetic charges transform as:
 \begin{align}
 &N^{f1}_m=N^a_m+ M,
N^{f2}_m=-N^a_m-N^b_m+M,   \nonumber\\
&N^{f3}_m=N^b_m-M,\nonumber
\end{align}
and electric charges transform as 
 \begin{align}
 &N_A=N^{f1}+N^{f2}-N^{f3},
  N_a=N^{f1}-N^{f2}, \nonumber\\
  &N_b=N^{f3}-N^{f2}.\nonumber
  \end{align}   
  For convenience, we can easily derive the following useful formulae:   
  \begin{align}
  &M=N^{f1}_m+N^{f2}_m+N^{f3}_m, 
   N^a_m=-N^{f2}_m-N^{f3}_m,\nonumber\\
&N^b_m=N^{f1}_m+N^{f2}_m+2N^{f3}_m.\nonumber
\end{align}
To summarize, a composite particle can be uniquely labeled by six numbers (three electric charges and three magnetic charges). The above relations can be recast in matrix form,
\begin{align}
&\left(\begin{matrix}
N_A\\
N_a\\
N_b
\end{matrix}\right)=\left(\begin{matrix}
1&1&-1\\
1&-1&0\\
0&-1&1\\
\end{matrix}\right)
\left(\begin{matrix}
N^{f1}\\
N^{f2}\\
N^{f3}
\end{matrix}\right)\,,\label{eqn:transition_matrix}
\\
&\left(\begin{matrix}
M\\
N_m^a\\
N_m^b
\end{matrix}\right)=\left(\begin{matrix}
1&1&1\\
0&-1&-1\\
1&1&2
\end{matrix}\right)
\left(\begin{matrix}
N^{f1}_m\\
N^{f2}_m\\
N^{f3}_m
\end{matrix}\right)\,,\label{eqn:transition_matrix1}
\end{align}
where the  two matrices belong to the $\mathbb{GL}(3,\Z)$ group.  
  All magnetic charges take values in an integral domain, i.e., $M$, $N_m^a$, $N_m^b$, $N_m^{fi}\in\Z$, where $i=1,2,3$. 
However, we will soon see that this integral domain will be potentially restricted to a smaller domain if we only consider the deconfined \emph{excitations} in the presence of a composite condensate.  We will introduce the notion of excitations in Sec.~\ref{section_of_screen}.

By the bare electric charge, we mean that $N_A$ is a naive count of the EM electric charge. In Sec.~\ref{section_of_screen}, it will be shown that   composite condensates will partially screen the charge, leading to a \emph{net} EM electric charge $Q$ to be defined in Eq.~(\ref{NE1}).
The electric charges $N^{fi}$ ($i=1,2,3$) are related to the number of attached fermions via the Witten effect formula:
\begin{align}
&N^{fi}=n^{fi}+\frac{\theta}{2\pi}N^{fi}_m\,,\,\text{with }n^{fi}\in\Z\,.\label{Nnf1}
\end{align}
The integer $n^{fi}$ counts the total number of fermions $f^i$ in the composite, and $\theta$ is determined by the  $\mathbb{Z}_2$ index of a 3D time-reversal invariant topological insulator. If $\theta=0$,   the partons occupy  a trivial band structure; if $\theta=\pi$, the partons occupy a  non-trivial topological insulator band structure.   
 The defining domains of $N^{f1,f2,f3},N_a,N_b,N_A$ can be either integer or potentially half-integer, depending on $\theta$.   

  \begin{figure}[t]
\centering
\includegraphics[width=8.5cm]{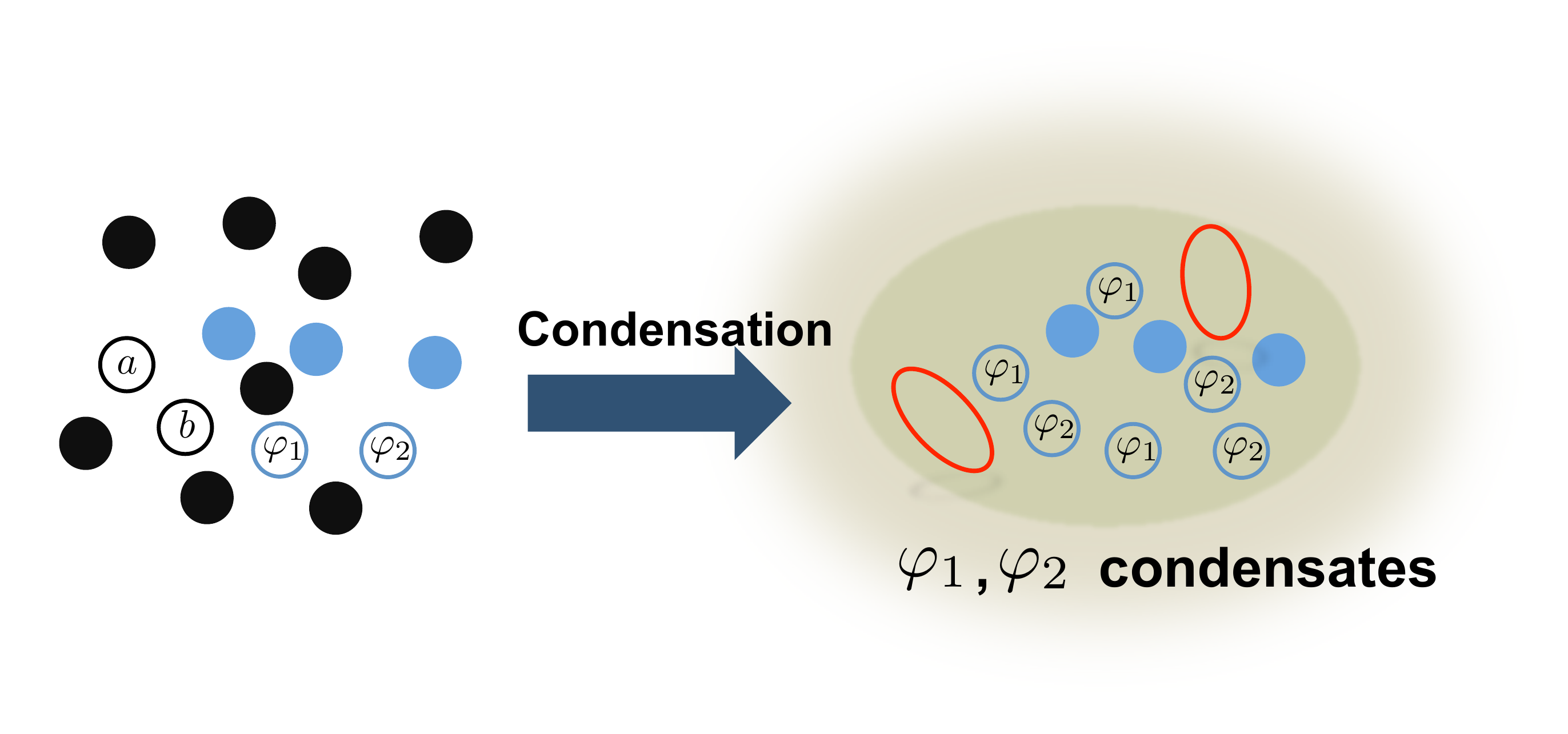}
\caption{(Color online) Schematic representation of composite particle condensation. Before condensation, the system is an electromagnetic plasma of composites in the $U(1)_\textrm{EM}\times U(1)_a\times U(1)_b$ gauge group. There are many composite particles (denoted by solid circles) including $\varphi_1$ and $\varphi_2$. There are also two gapless photons (denoted by $a$ and $b$ in the figure), indicating that the phase before condensation is a gapless Coulomb phase for both internal dynamical gauge fields. After condensing $\varphi_1$ and $\varphi_2$, the system enters a gapped phase in the absence of photons. All composites (denoted by black solid circles on the left) that have nonzero mutual statistics with both $\varphi_1$ and $\varphi_2$ are confined. Otherwise, those composites that have trivial (zero) mutual statistics with both condensates survive  as excitations (denoted by blue solid circles) of the gapped phase. The red loops on the right represent loop excitations due to the two condensates.}
\label{figure_condensation}
\end{figure}

\begin{table*}[]
\centering
\caption{The electric charges and magnetic charges of composite condensates $\varphi_1$ and $\varphi_2$ are determined by parameters $(u,v,u',v',q,q',\theta)$. The charges are shown for both the $U(1)_a\times U(1)_b\times U(1)_{\rm EM}$ gauge group labels and the $U(1)_{f1}\times U(1)_{f2}\times U(1)_{f3}$ gauge group labels, both of which are used interchangeably in the main text. The two sets of parameters are related via    Eqs.~(\ref{eqn:transition_matrix},\ref{eqn:transition_matrix1}).  The value of  $\theta=0,\pi$ is determined by the band structure topology of the partons. Given $\theta$, the other six parameters are constrained by several conditions listed in Eqs.~(\ref{constr_1plus},\ref{constr_1},\ref{constr_3},\ref{constr_5}). A concrete example of composite condensates that generate a fractional topological insulators (FTI) with $\Theta=\frac{\pi}{9}$  discussed  in  Sec.~\ref{sec_FTI} is also shown. For this case, we see that $\varphi_1$ is a bosonic bound state of two $a_\mu$ magnetic monopoles,  two $b_\mu$ magnetic monopoles, one $f^1$ parton,  four $f^2$ partons, and  one $f^3$ parton, while $\varphi_2$ is a bosonic bound state of four $a_\mu$ magnetic monopoles, ten $b_\mu$ magnetic monopoles,   nine $f^2$ partons, and three hole-like $f^3$ partons. We may call the partons themselves as the simplest  composites although literally they are not composites. Also, the electron is just a collection of one $f^1$, one $f^2$, and one $f^3$.}
\label{table:em}
\resizebox{\textwidth}{!}{%
\begin{tabular}{c|cccccc|  ccccccccc}
\hline

\hline

\hline
\hline
\multirow{2}{*}{\textbf{Composite particles}} & \multicolumn{6}{c|}{$U(1)_a\times U(1)_b\times U(1)_{\rm EM}$} & \multicolumn{9}{c}{$U(1)_{f1}\times U(1)_{f2}\times U(1)_{f3}$}                                                                                      \\ \cline{2-16} 
        \textbf{~}         & $N_a$     & $N_b$     & $N_A$     & $N_m^a$     & $N_m^b$     & $M$    & $N^{f1}$ & $N^{f2}$ & $N^{f3}$ & $N^{f1}_m$ & $N^{f2}_m$ & $N^{f3}_m$ & $n^{f1}$                   & $n^{f2}$                        & $n^{f3}$                   \\ \hline
$\varphi_1$       & $0$       & $0$       & $q$       & $u$         & $v$         & $0$    & $q$      & $q$      & $q$      & $u$        & $-u-v$     & $v$        & $q-\frac{\theta}{2\pi}u$   & $q+\frac{\theta}{2\pi}(u+v)$    & $q-\frac{\theta}{2\pi}v$   \\ \hline
$\varphi_2$       & $0$       & $0$       & $q'$      & $u'$        & $v'$        & $0$    & $q'$     & $q'$     & $q'$     & $u'$       & $-u'-v'$   & $v'$       & $q'-\frac{\theta}{2\pi}u'$ & $q'+\frac{\theta}{2\pi}(u'+v')$ & $q'-\frac{\theta}{2\pi}v'$ \\  \hline

 \hline
$\varphi_1$ of {FTI} in Sec.~\ref{sec_FTI}      & $0$       & $0$       & $2$      & $2$        & $2$        & $0$    & $2$     & $2$     & $2$     & $2$       & $-4$   & $2$       & $1$ & $4$ & $1$ \\  \hline
$\varphi_2$  of {FTI} in Sec.~\ref{sec_FTI}        & $0$       & $0$       & $2$      & $4$        & $10$        & $0$    & $2$     & $2$     & $2$     & $4$       & $-14$   & $10$       & $0$ & $9$ & $-3$ \\ \hline 
Parton $f^1$     & $1$       & $0$       & $1$      & $0$        & $0$        & $0$    & $1$     & $0$     & $0$     & $0$       & $0$   & $0$       & $1$ & $0$ & $0$  \\ \hline
Parton $f^2$     & $-1$       & $-1$       & $1$      & $0$        & $0$        & $0$    & $0$     & $1$     & $0$     & $0$       & $0$   & $0$       & $0$ & $1$ & $0$  \\ \hline
Parton $f^3$     & $0$       & $1$       & $-1$      & $0$        & $0$        & $0$    & $0$     & $0$     & $1$     & $0$       & $0$   & $0$       & $0$ & $0$ & $1$ \\ \hline
 Electron     & $0$       & $0$       & $1$      & $0$        & $0$        & $0$    & $1$     & $1$     & $1$     & $0$       & $0$   & $0$       & $1$ & $1$ & $1$  \\ \hline\hline

\hline

\hline
\end{tabular}
}
\end{table*}
\subsection{A local field theoretic description of condensed composites}
Since there are two internal gauge fields with strong gauge fluctuations, we can  consider  two   linearly independent Bose condensates denoted by $\varphi_1$  and  $\varphi_2$, as shown in Fig.~\ref{figure_condensation}. Both condensates should contain  magnetic monopoles of the internal gauge fields but be  neutral under  both the   $U(1)_a$ and $U(1)_b$ gauge groups, i.e., $N_a=0\,,N_b=0$. Since the EM gauge field is treated as a background gauge field for the purpose of the EM response, the condensates should not carry $M$. Otherwise, the EM gauge field must be strongly fluctuating, which is not our working assumption.  
In summary, the electric and magnetic charges of $\varphi_1$ and $\varphi_2$ can be completely determined by six parameters $(q,u,v,q',u',v')$ in Table~\ref{table:em}. Since the condensates are not dyonic in each gauge group the order parameters $\langle \varphi_1\rangle$ and $\langle \varphi_2\rangle$ are local to each other and can be described by an effective local quantum  field theory.  More concretely, we may start with a phenomenological Ginzburg-Landau-type action in   4D Euclidean spacetime: 
\begin{align}
\!  \! \!  \!\! S_{\rm GL}\!=\!\int \!d^4x \sum^2_I\!\left(|\hat{D}_\mu\varphi_I |^2+\mu^2 |\varphi_I|^2\! +\! \lambda |\varphi_I|^4\right)+ \!S_M,\label{equation_free_energy}
\end{align}
where  the Ginzburg-Landau parameter $\lambda$ is positive. $\hat{D}_\mu$ is the covariant derivative defined by: 
\begin{align}
\hat{D}_\mu\equiv \partial_\mu+i  N_A A_\mu+i  N_m^a\tilde{a}_\mu+i  N_m^b \tilde{b}_\mu.
\end{align}
Here, $N_A,N^a_m,N^b_m$ are two sets of electric/magnetic charges of $\varphi_1$ and $\varphi_2$, which can be found in Table~\ref{table:em}. The one-form gauge fields $\tilde{a}_\mu$ and $\tilde{b}_\mu$ serve as the  magnetic dual of the gauge fields $a_\mu$ and $b_\mu$, respectively.  For example, $\tilde{a}_\mu$ is introduced such that its gauge charge is carried by magnetic monopoles of the  $U(1)_a$ gauge group. Meanwhile, the  magnetic flux of $\tilde{a}_\mu$ gives the electric field $\mathbf{E}^a$, namely, $\mathbf{E}^a=\nabla\times \mathbf{\tilde{a}}$. $\tilde{b}_\mu$ can be understood analogously to $\tilde{a}_\mu$. $S_M$ includes the Maxwell terms:
$S_M=\int d^4x(\frac{1}{4}\tilde{f}^a_{\mu\nu}\tilde{f}^a_{\mu\nu}+\frac{1}{4}\tilde{f}^b_{\mu\nu}\tilde{f}^b_{\mu\nu})\,.$ 
In the condensed phase where the mass parameter $\mu^2<0$,    nonzero expectation values $\langle\varphi_I\rangle\neq0$ develop.
Here, $\tilde{f}^{a,b}_{\mu\nu}$ are the field strength tensors of $\tilde{a}_\mu$ and $\tilde{b}_\mu$. One advantage to  using dual gauge fields is that the problem of strong gauge fluctuations ($g_a\gg 1\,,g_b\gg 1$) of the $a_\mu$ and $b_\mu$ gauge fields is transformed into the problem of weak gauge fluctuations of the dual gauge fields $\tilde{a}_\mu$ and $\tilde{b}_\mu$ by noting that the coupling constants between magnetic charges and dual gauge fields is the inverse of the original coupling constants, i.e., $1/g_{a,b}$.

It is noteworthy that the six numbers $(u,v,u',v',q,q')$ describing the condensates are not completely free since the following three conditions should be satisfied:
 \begin{enumerate}
\item $\varphi_1$ and $\varphi_2$ are bosonic;\label{condition_phi2}
\item $\varphi_1$ and $\varphi_2$  are allowed to condense simultaneously;\label{condition_phi3}
\item $\varphi_1$ and $\varphi_2$ are linearly independent;\label{condition_phi1}
\end{enumerate} 
 such that the composite condensates $\varphi_1$ and $\varphi_2$ are physically viable.  In more detail, according to the   domains of definition of every charge (e.g., all magnetic charges are integer-valued, all $n^{fi}$ are integer-valued), we  can deduce the domains of the six numbers (see Table~\ref{table:em}):
   \begin{align}
&u\,,v\,,u'\,,v'\,,q-\frac{\theta}{2\pi}u \,,  q+\frac{\theta}{2\pi}(u+v) \,,q-\frac{\theta}{2\pi}v \in\Z,\label{constr_1plus}\\
& q'-\frac{\theta}{2\pi}u' \,,q'+\frac{\theta}{2\pi}(u'+v')\, ,q'-\frac{\theta}{2\pi}v'   \in\Z\,.\label{constr_1}
\end{align}
Since only bosonic particles can undergo Bose  condensation, one should carefully check the self-statistics of $\varphi_1$ and $\varphi_2$. Furthermore, the mutual statistics between $\varphi_1$ and $\varphi_2$ must be zero so that they are allowed to condense simultaneously. 
 
Let us first consider the latter. The trivial mutual statistics   between two composites (with and without prime) is given by the following equation:
\begin{align}
 \sum^3_{i}(N^{fi}_m{n^{fi}}'-{N^{fi}_m}'n^{fi})=0\label{equation:mutual_stat}
\end{align}
or equivalently: $\sum^3_{i}(N^{fi}_m{N^{fi}}'-{N^{fi}_m}'N^{fi})=0\,.$ 
 If this equation is satisfied, then the two composites can   condense simultaneously. Furthermore, condensation of one of the composites will lead to deconfined particles (an excitation spectrum)  having electric and magnetic charges are determined by this equation.  
If the equation is not satisfied, the condensation of one of the composites will confine the other \cite{cardy1,cardy2}.
 Inserting the electric and magnetic charges of $\varphi_1,\varphi_2$ into the equation,  it turns out that $\varphi_1$ and $\varphi_2$ always satisfy the condition  of trivial mutual statistics.

Next, we need to further check the self-statistics of $\varphi_1$ and $\varphi_2$. For a generic composite, the self-statistics phase $e^{i\pi \Gamma}$ is determined by the following integer:
\begin{align}
\Gamma=\sum_i^3 (N_m^{fi}n^{fi}+n^{fi})\,,\label{eq:quantum_stat}
\end{align}
where the second term $n^{fi}$ counts the number of fermionic partons inside the composite.  The first term $N_m^{fi}n^{fi}$ arises from the angular momentum of the relative motion between the electric  charge and magnetic charge.  Note  that the polarization charge ``$\frac{\theta}{2\pi}N^{fi}_{m}$''  due to the Witten effect in Eq.~(\ref{Nnf1}) does not enter the statistics. A field theoretic understanding of this phenomenon can be found in Ref.~\onlinecite{goldhaber89}. 
For later convenience, we may express Eq.~(\ref{eq:quantum_stat}) in terms of the $U(1)_{\rm EM}\times U(1)_a\times U(1)_b$ gauge groups: 
\begin{align}
\Gamma=&N^a_m(n^{f1}-n^{f2})+N^b_m (n^{f3}-n^{f2})\nonumber\\
&+(M+1)(n^{f1}+n^{f2}-n^{f3})\,,\label{eq:stat123}
\end{align}
where we have added even integers during the derivation as only the value of $\Gamma$ mod $2$ is meaningful. 
If $\Gamma$ is even, the composite is bosonic; otherwise, it is fermionic. After inserting the values of the electric and magnetic charges of $\varphi_1$ and $\varphi_2$ into $\Gamma$, we may obtain the $\Gamma$ formulae of both $\varphi_1$ and $\varphi_2$ (denoted as $\Gamma(\varphi_1),\Gamma(\varphi_2)$) as functions of $u,v,u',v',q,q'$ (see Appendix~\ref{appendix_bosonic}). The requirement that both $\varphi_1$ and $\varphi_2$ are bosonic leads to the following constraints on $u,v,u',v',q,q'$:
 \begin{align}
&\Gamma(\varphi_1)\in\Z_{\rm even}\,,\,~~\Gamma(\varphi_2)\in\Z_{\rm even}\,.\label{constr_3}
\end{align}
So far, we have deduced several constraints on the six numbers: Eqs.~(\ref{constr_1plus},\ref{constr_1},\ref{constr_3}), 
but there is one more constraint, i.e., Eq. (\ref{constr_5}), which enforces that $\varphi_1$ and $\varphi_2$ are linearly independent.
It is possible that one of the composites consists of   several copies of the other composite, in which case there is actually only one condensate.  To avoid this situation, the  following condition should be strictly imposed:
  \begin{align}
 uv'-u'v\neq 0\,.\label{constr_5}
 \end{align}
 A physical understanding of this condition will be presented in Sec.~\ref{sec:loop_loop}. 
 For convenience, we introduce the following notation:
\begin{align}
&K=\bpm u & v \\ u'& v'\epm,~~ \mathbf{N}_m=\bpm N_m^a\\N_m^b\epm,~~ \mathbf{q}=\bpm q\\q'\epm,\label{define_KNq1}
\\
&\mathbf{\Phi}_e=\bpm \Phi^a_e\\ \Phi^b_e\epm, ~~\mathbf{N}_e=\bpm N_a\\N_b\epm.
\label{define_KNq}
\end{align}
Then, the matrix $K$ is invertible, namely, its determinant should be nonzero, as given by Eq.~(\ref{constr_5}). 
In summary,  the conditions Eqs.~(\ref{constr_1plus},\ref{constr_1},\ref{constr_3},\ref{constr_5}) should be imposed on the six integers $u,v,q;u',v',q'$ such that the two condensates satisfy conditions: (\ref{condition_phi2},\ref{condition_phi3},\ref{condition_phi1}). 

  \subsection{Generalized flux quantization and   loop excitations}\label{sec:loop_loop}

In order to gain a better physical understanding of the condition (\ref{constr_5}), we need to carefully study the ``generalized flux quantization'' induced by the two condensates $\varphi_1$ and $\varphi_2$ whose electric and magnetic charges are listed in Table~\ref{table:em}.  In a usual type-II  superconductor, we know that the EM magnetic flux denoted by $\Phi^A_M$ is screened and quantized according to $2\Phi^A_M/2\pi=\Phi^A_M/\pi\in\Z$ since the Cooper pair condensate carries $2e$ EM electric charge.  In our case, the  two condensates $\varphi_1$ and $\varphi_2$ carry not only  EM electric charges but also   magnetic charges of the $a$ and $b$ gauge groups as shown in Table~\ref{table:em}. As a result, we have the following generalized flux quantization conditions:
\begin{align}
& q\Phi^A_M+u\Phi^a_e+v\Phi^b_e=2\pi\ell,\label{EM1}\\
&q'{\Phi^A_M}+u'{\Phi^a_e}+v'{\Phi^b_e}=2\pi\ell',\!\label{EM1+}
\end{align}
where $\Phi^A_M$ is the EM magnetic flux piercing a spatial loop $\mathcal{S}^1$. $\Phi^a_e$ and $\Phi^b_e$ are the $a$- and $b$-electric fluxes piercing $\mathcal{S}^1$, respectively.  Here, instead of magnetic fluxes,  electric fluxes of  the $U(1)_a\times U(1)_b$ gauge group are involved since the condensates carry magnetic charges rather than electric charges of the $U(1)_a\times U(1)_b$  gauge group. $\ell,\ell'\in\Z$ label the winding numbers of the mapping $\mathcal{S}^1\rightarrow U(1)$ of the condensate order parameters $\varphi_1,\varphi_2$.   

 In contrast with fluxes of the internal gauge groups, arbitrary  values of $\Phi^A_M$ are allowed to be inserted. In other words, $A_\mu$ itself is not higgsed, and  the EM electric charge of the electrons  is a well-defined quantum number. This implies that the two condensates must provide a new charge screening mechanism such that the   net EM electric charge of each condensate is zero, although both condensates carry a nonzero bare EM electric charge ($N_A=q,q'$).  This screening effect can lead to fractionalization of the charge of excitations, even in the absence of an external EM magnetic charge. We will postpone a discussion of this issue  until Sec.~\ref{section_of_screen}.

 Since $A_\mu$ is an external non-dynamical field, we may temporarily turn it off in Eqs.~(\ref{EM1},\ref{EM1+}) to find:
\begin{align}
&  u\Phi^a_e+v\Phi^b_e=2\pi \ell\,, \,u'{\Phi^a_e}+v'{\Phi^b_e}=2\pi \ell'\,.\label{phie1}
\end{align}
 A generic solution of Eq.~(\ref{phie1}) is given by:
 \begin{align}
&\Phi^a_e=2\pi\frac{\ell v'-\ell'v}{\mathsf{Det}K}\,,\,\Phi^b_e=2\pi\frac{\ell'u-\ell u'}{\mathsf{Det}K}\,.\label{phi_ae}
 \end{align}
 Here, $u,v,u',v'\in\Z$ satisfy the condition (\ref{constr_5}). 
 By noting that $\ell v'-\ell'v$ is divisible by  the greatest common divisor  ${\mathsf{GCD}}(v,v')$ and $\ell'u-\ell u'$ is divisible by the greatest common divisor $\mathsf{GCD}(u,u')$, one can use  {B\'{e}zout's lemma} (see Appendix \ref{appendix_proof_theorem_flux}) to obtain
 the minimal quantized electric fluxes:  
   \begin{align}
\!\!\!\!(\Phi^a_e)_{\rm min}\!=\!2\pi\!\bigg|\frac{ \mathsf{GCD}(v,v')}{ \mathsf{Det}K}\bigg|, (\Phi^b_e)_{\rm min}\!=\!2\pi\!\bigg|\frac{ \mathsf{GCD}(u,u')}{ \mathsf{Det}K}\bigg|.\!\label{discreteflux1}
 \end{align}
Since $| uv'-u'v|$ is divisible by both $\mathsf{GCD}(v,v')$ and $\mathsf{GCD}(u,u')$, we have the following two useful inequalities: 
\begin{align}
\!\!\!\!\!|uv'-u'v|\!\geq\!|\mathsf{GCD}(u,u')|\,, |uv'-u'v|\!\geq\!|\mathsf{GCD}(v,v')|.\label{eq:ineq}
\end{align}

Based on B\'{e}zout's lemma, we can easily prove the following theorem. The proof is shown in Appendix \ref{appendix_proof_theorem_flux}:
\begin{thm}
  $(\Phi^a_e)_{\rm min}=2\pi$ and $(\Phi^b_e)_{\rm min}=2\pi$ if and only if   $|uv'-u'v|=1$.\label{theorem_flux}
\end{thm}
The above theorem leads to the following criterion for loop/flux excitations:
\begin{crt}
[Criterion for loop excitations]  If $|\mathsf{Det}K|=1$, the bulk has no deconfined discrete gauge fluxes (therefore, no detectable loop excitations); If $|\mathsf{Det}K|>1$, the bulk has deconfined discrete gauge fluxes (therefore, detectable loop excitations with minimal flux strength smaller than $2\pi$.)\label{crt_loop_exc}
\end{crt}

The solutions $(\Phi_e^a, \Phi_e^b)$  in Eq.~(\ref{phi_ae}) can be recast in the following form:
\begin{align}
\mathbf{\Phi}_e=2\pi K^{-1}\mathbf{L}  \,,\label{eq:phie_L}
\end{align}
where the integer vector $\mathbf{L}=(\ell,\ell')^T$. Thus,  we may define a 2D loop-lattice generated by a dimensionless integer vector $\mathbf{L}$:
\begin{dfn}
 [{Loop-lattice}] A loop-lattice is a 2D square lattice where each site  corresponds to a loop excitation labeled by $\mathbf{L}=(\ell,\ell')^T$. The corresponding electric flux strength $\mathbf{\Phi}_e$ of each site is determined by Eq.~(\ref{eq:phie_L}).\label{dfn_loops}
\end{dfn}

    \subsection{Point-particle excitations and charge fractionalization}\label{section_of_screen}
In addition to loop excitations, we also have point-particle excitations: 
  \begin{dfn}
 [{Excitation} and {Charge lattice}] Excitations are defined as deconfined particles that have trivial mutual statistics with both condensates. All excitations form a 4D charge lattice which is a sublattice of the original 6D lattice. Unless otherwise specified, excitations always refer to point-particle excitations.\label{dfn_excitation}
\end{dfn}
 By definition, all excitations have trivial mutual statistics with respect to the condensates. In other words, Eq.~(\ref{equation:mutual_stat}) holds between any excitation and $\varphi_1$, and also holds between any excitation and $\varphi_2$. By explicitly using the parameters of $\varphi_1$ and $\varphi_2$ in Table~\ref{table:em}, the electric and magnetic charges of excitations are constrained by the following two equations:
   \begin{align}
&qM=uN_a+vN_b\,, ~~q'M=u'N_a+v'N_b\,.\label{excitation_1}
   \end{align}
   Therefore, a generic particle that has six independent charges $(N_A,N_a,N_b,M,N_m^a,N_m^b)$ is now completely determined by four of them $(N_A,M,N_m^a,N_m^b)$ if the particle is a deconfined excitation in the condensed phase.   
 Keeping Eq.~(\ref{constr_5}) in mind, $N_a$ and $N_b$ are fully determined by $M$: 
$ N_a=\frac{qv'-q'v}{\mathsf{Det}K}M\,,\, N_b=\frac{q'u-qu'}{\mathsf{Det}K}M\,$
 which can be written as  
 \begin{align}
 \mathbf{N}_e=MK^{-1}\mathbf{q}\label{NabM}
 \end{align}
 by using the notation in Eqs.~(\ref{define_KNq1},\ref{define_KNq}).
 
  \begin{figure}[t]
\centering
\includegraphics[width=8cm]{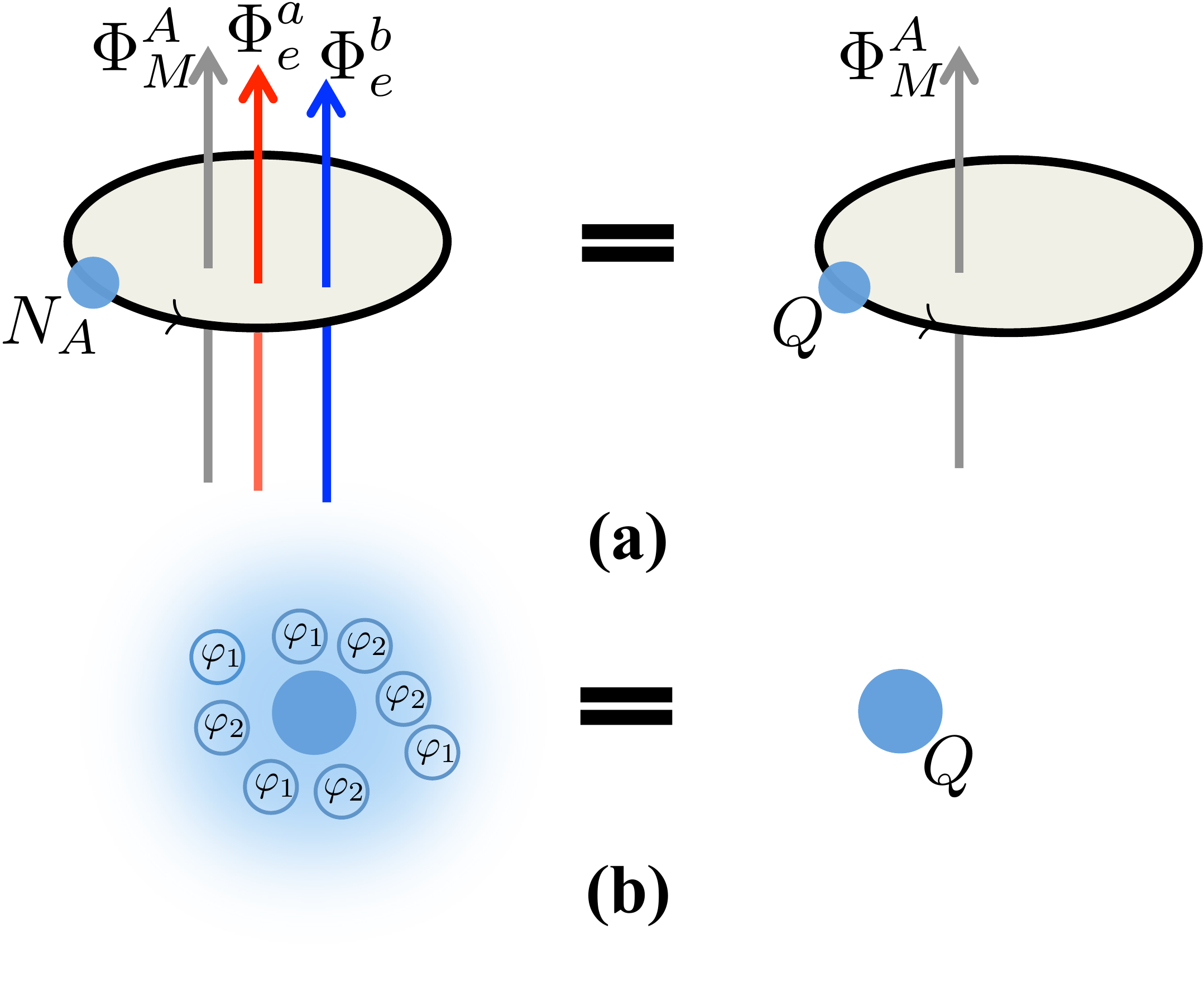}
\caption{(Color online) Schematic representation of the charge screening mechanism. Consider a composite particle carrying $N_A$ units of the EM ($A_\mu$) electric charge (i.e., $U(1)$ symmetry charge), $N_m^a$ units of magnetic charge of the $a_\mu$ field, and $N_m^b$ units of magnetic charge of the $b_\mu$ gauge field.  Due to the condensates, $N_A$ is partially screened such that the net EM electric charge $Q$ is given by Eq.~(\ref{NE1}), which is different from $N_A$. In (a) the physics of Aharonov-Bohm  effect in Eq.~(\ref{ABPhase}) is illustrated. An excitation (denoted by the blue ball) adiabatically moves along a closed trajectory and feels the EM magnetic flux $\Phi^A_M$, the electric flux $\Phi^a_e$ of the $a_\mu$ gauge field, and the electric flux of the $b_\mu$ gauge field. In (b), condensed particles $\varphi_1$ and $\varphi_2$ form a Debye-H{\"u}ckel-like charge cloud around an excitation, providing the screening charge $Q_{\rm Debye}$ in Eq.~(\ref{NE_screen}).}
\label{figure_flux}
\end{figure}

As mentioned in Sec.~\ref{sec:gauge_strc}, the EM electric charge of a particle, $N_A$, is called the ``bare'' charge, which suggests that it will be partially screened due to the condensates. In order to clearly  see  the screening, we turn on the external EM field $A_\mu$ to probe the EM response and  consider a spatial loop $C$.  The total Aharonov-Bohm  phase accumulated by an adiabatically moving test particle is given by (see Fig.~\ref{figure_flux}):
\begin{align}
&\text{Aharonov-Bohm phase}\nonumber\\
=
&\exp\{iN_A \Phi^A_M+i M \Phi^A_E+iN^a\Phi^a_m+iN^b\Phi^b_m\nonumber\\
&+iN^a_m \Phi^a_e+iN^b_m \Phi^b_e\}\,,\label{ABPhase}
\end{align}
where $\Phi^A_E$ is EM electric flux piercing $C$, $\Phi_m^a$ and $\Phi_m^b$ are the $a$- and $b$-magnetic fluxes respectively.  However,  Eqs.~(\ref{EM1},\ref{EM1+}) indicate that $\Phi^a_e$ and $\Phi^b_e$ depend linearly   on the   external EM magnetic flux $\Phi^A_M$.  Solving Eqs.~(\ref{EM1},\ref{EM1+}) leads to:
 \begin{align}
&\Phi^a_e=2\pi\frac{kv'-k'v}{\mathsf{Det}K}-{ \Phi^A_M\frac{qv'-q'v}{\mathsf{Det}K}}\,,\label{screening1}\\
&\Phi^b_e=2\pi\frac{k'u-ku'}{\mathsf{Det}K}-{ \Phi^A_M\frac{q'u-qu'}{\mathsf{Det}K}}\,.\label{screening2}
 \end{align}
The terms that depend linearly on $\Phi_M^A$  correct the saddle point solutions in Eq.~(\ref{phi_ae}). 
 
Taking Eqs.~(\ref{screening1},\ref{screening2}) into account, the contribution to the Aharonov-Bohm phase due to the external EM gauge field can be isolated. Eq.~(\ref{ABPhase}) can be recast into  $e^{iQ \Phi^A_M+\cdots}$,   where $\cdots$ denotes the remaining terms that do not contain the factor $\Phi^A_M$, and, $Q$ is the \emph{net EM electric charge}:
  \begin{align}
 Q=N_A-Q_{\rm Debye}\,, \label{NE1}
 \end{align}
 where 
  \begin{align}
\!\!\!\! Q_{\rm Debye}\!=\!{ N_m^a\frac{qv'-q'v}{\mathsf{Det}K}\!+\!N_m^b\frac{q'u-qu'}{\mathsf{Det}K}}\!=\!\mathbf{N}_m^T K^{-1}\!\mathbf{q}\,.\!\!\!\label{NE_screen}
 \end{align}
   denotes the charge carried by the Debye-H{\"u}ckel-like  screening   cloud (see Fig.~\ref{figure_flux}). The matrix $K$, vector $\mathbf{N}_m,$ and the vector $\mathbf{q}$ are defined in Eq.~(\ref{define_KNq}).   
Concerning the screening charge, there are two interesting limits. First,  the total EM electric charge of each condensate is zero, i.e.,  $Q=0$ for both condensates $\varphi_1$ and $\varphi_2$ (see Table~\ref{table:em}). For example, we have $Q_{\rm Debye}=q$ for $\varphi_1$, which completely screens its bare EM electric charge $N_A=q$. 
Second,  let us consider an intrinsic excitation whose bare EM electric charge vanishes, $N_A=0$. Its  net EM electric charge $Q$ is nonzero and completely given by that of the  Debye screening cloud: 
$
Q=-Q_{\rm Debye}= - \mathbf{N}_m^T K^{-1}\!\mathbf{q}
$.

In fact,   charge fractionalization in   Abelian FQH states can also be understood via the above Aharonov-Bohm thought experiment. As an example, let us derive the   fractionalization of charge  in  the $\nu=1/3$ Laughlin state. The effective field theory is described by the following Lagrangian:
 \begin{align}
  \mathcal{L}=\frac{3}{4\pi}a_\mu\partial_\nu a_\lambda \epsilon^{\mu\nu\lambda}+\frac{1}{2\pi}A_\mu \partial_\nu a_\lambda\epsilon^{\mu\nu\lambda}\,.
    \end{align} 
Here the gauge field $a_\mu$ is a dual description of the electron current $J_\mu$:  $J_\mu=\frac{1}{2\pi}\partial_\nu a_\lambda \epsilon^{\mu\nu\lambda}$. The second term in the Lagrangian $\mathcal{L}$ means that each electron carries one unit of electric charge. Excitations in   FQH states are labeled by gauge charges of the $a_\mu$ gauge group, since they minimally couple to $a_\mu$. In this sense, let us consider the Aharonov-Bohm experiment for an excitation that carries one unit of gauge charge of $a_\mu$. The Aharonov-Bohm phase is given by $e^{i\Phi_a}$ where $\Phi_a$ is the $a_\mu$ magnetic flux felt by the excitation. In the hydrodynamical field theory $\mathcal{L}$, $\frac{1}{2\pi}\Phi_a$ corresponds to the electron  density. By studying the equation of motion of $a_\mu$ in $\mathcal{L}$, we obtain: $\Phi_A=3\Phi_a$. Physically, this identity means that each electron effectively corresponds to three units of magnetic flux   of the background EM field, which is nothing but the definition of  filling fraction $\nu=\frac{1}{3}$. Thus, the Aharonov-Bohm phase accumulated by the excitation is identical to: $e^{i\Phi_a}=e^{i\frac{1}{3}\Phi_A}$. The coefficient $\frac{1}{3}$ indicates that the excitation in the presence of $A_\mu$ behaves as an electrically charged particle with $\frac{1}{3}$ charge.
   
Let us come back to our 3D system. We introduce the following  two equivalent criteria for charge fractionalization:
  \begin{crt}
[Criterion for charge fractionalization]  Charge fractionalization exists if excitations with zero $M$ and fractionalized $Q$ exist. \label{crt_charge_1}
 \end{crt}  
and,
 \begin{crt}
[Criterion for charge fractionalization]  Equivalently,  charge fractionalization  exists if  the EM magnetic charge  $M$ of excitations is quantized in units of an integer $w>1$, i.e., $M=0,\pm w, \pm 2w,\cdots$. \label{crt_charge_2}
 \end{crt}   
 In Appendix \ref{appendix_dirac}, the equivalence of the above two criteria is explained by using the well-known Dirac-Zwanziger-Schwinger quantization condition. 
 The requirement of $M=0$ in Criterion~\ref{crt_charge_1} can be understood as follows. Typically, in the presence of $M$, excitations can potentially carry a fractionalized $Q$ due to the Witten effect. However, this does not mean our 3D quantum system is fractionalized. The topological insulator (TI) is a typical example. If a single EM monopole ($M=1$) is inserted into the bulk, there is a half-charge cloud surrounding the monopole  \cite{Qi2008,franz}. However, since  the TI can be  realized in a non-interacting band insulator we do not consider it to be fractionalized. 
In order to highlight the set of excitations with $M=0$, we  introduce the notion of  intrinsic excitations and  intrinsic charge lattice:
    \begin{dfn}
[{Intrinsic excitations} and {intrinsic charge lattice}] Intrinsic excitations are excitations with zero EM magnetic charge, i.e., $M=0$;
the intrinsic charge lattice is  a special 3D charge lattice with zero EM magnetic charge, i.e., $M=0$.\label{dfn_intrinsic_exc}
 \end{dfn}

One can verify that   $Q_{\rm Debye}$ in Eq.~(\ref{NE1}) is the \emph{unique} source of   charge fractionalization. In other words, $N_A$ in Eq.~(\ref{NE1}) is always integer-valued when $M=0$ (see Appendix \ref{appendix_na_integer} for details);   charge fractionalization exists if and only if $Q_{\rm Debye}$ is fractional when $M=0$.

Finally, we   show that the  Debye charge cloud $Q_{\rm Debye}$ in Eq.~(\ref{NE_screen}) can also be understood in a more formal way, i.e., from a topological $BF$ field theory. 
 Without loss of generality, we consider the London limit (i.e., deep in the confined phase) such that the amplitude fluctuations of $|\varphi_I|$ are negligible. In this limit, we may dualize $S_{\rm GL}$ in Eq.~(\ref{equation_free_energy}) into a two-component topological $BF$ field theory \cite{horowitz89}:
\begin{align}
S=&i\int \frac{1}{2\pi} \mathscr{B}^T  \wedge  K d \mathscr{A}+i\int \frac{1}{2\pi} \mathbf{q}^T \mathscr{B}\wedge d A+\mathcal{S}_{ex},\label{BFaction}
\end{align}
where we define the two-component vectors $\mathscr{B}=\bpm \mathcal{B},\mathcal{B}'\epm^T$ and   $\mathscr{A}=\bpm \tilde{a},\tilde{b}\epm^T$, and use a differential form notation. Here, $\mathcal{B}$ and $\mathcal{B}'$ are two Kalb-Ramond $2$-form gauge fields introduced as a result of the particle-vortex line duality transformation in (3+1)D \cite{savit}. Physically, they are related to the supercurrents of the condensates $\varphi_1$ and $\varphi_2$, respectively, via: 
$J^{\varphi_1}=\frac{1}{2\pi} \star  d\mathcal{B}\,,\,{J^{\varphi_2}}=\frac{1}{2\pi} \star  d \mathcal{B}'\,,
$ where $\star$ is the usual Hodge-dual operation. Since the energy gap in the bulk of the topological $BF$ field theory is effectively infinite, the term $\mathcal{S}_{ex}$ is added by hand in order to take  into account  the  point-like excitations labeled by $\mathbf{N}_m=(N_m^a,N_m^b)^T,$ and the loop excitations labeled by the integer vector $\mathbf{L}=(\ell,\ell')^T$ (see Definition~\ref{dfn_loops}):
\begin{align}
\mathcal{S}_{ex}=i \int \mathbf{N}_m^T \mathscr{A} \wedge \star  j+i \int\mathbf{L}^T \mathscr{B} \wedge \star  \Sigma\,,\label{BFex}
\end{align}
where the vector $j$ denotes the composite excitation current, and  the tensor $\Sigma$ denotes the loop excitation current. Integrating out  the dynamical fields $\mathscr{A}$ and $\mathscr{B}$ yields an effective theory for $j$ and $\Sigma$ in the presence of the external EM field $A_\mu$:
\begin{align}
\!\!\!\!S_{\rm eff}\!= \!i\mathbf{N}_m^T K^{-1}\!\mathbf{q}  \!\!\int \! j\!\wedge \star A +\!i  2\pi\mathbf{N}_m^TK^{-1}\mathbf{L} \! \!\int \!\Sigma \wedge d^{-1}\,j.\!\! \label{eqn:effective}
\end{align}
It is   remarkable that the first term in the effective action (\ref{eqn:effective}) is nothing but the Debye screening charge cloud $Q_{\rm Debye}$ defined in Eq.~(\ref{NE_screen}). Thus, $Q_{\rm Debye}$ is a topological property of an excitation. The second term represents the long-range Aharonov-Bohm statistical interaction between fluxes and particles. The operator $d^{-1}$ is a formal notation defined as the operator inverse  of $d$, whose exact form can be understood in momentum space by Fourier transformation.  The coefficient $\mathbf{N}^T_m K^{-1} \mathbf{L}$ gives rise to the  charge-loop braiding statistics $\vartheta^{cl}$ between composite particles with quantum number $\mathbf{N}_m$ and loop excitations with electric fluxes $\mathbf{\Phi}_e=2\pi(K)^{-1}\mathbf{L}$ due to Eq.~(\ref{eq:phie_L}):
\begin{align}
\vartheta^{cl}=2\pi\mathbf{N}^T_m K^{-1} \mathbf{L}=\mathbf{N}^T_m  \mathbf{\Phi}_e \,.\label{eq:cltheta}
\end{align}

Now that we have carefully developed a theory that describes topological phases in the presence of $U(1)$ composite condensates we will use the results to construct fractionalized 3D topological insulators with time-reversal symmetry.


 \section{Fractional topological insulators}\label{sec:theta}
  In this section, we will study 3D topological phases of matter with non-vanishing axion  angle $\Theta$. The presence of nontrivial values of $\Theta$ lead to several observable phenomena including a surface quantum Hall effect, and the celebrated Witten effect: a magnetic monopole will bind a electric charge. For free-fermion time-reversal invariant topological insulators, the angle is $\pi \text{ mod } 2\pi$ \cite{Qi2008}.   In  fractionalized states where strong interactions and correlations are taken into account,  in principle, the axion angle can be fractional (i.e., $\Theta/\pi$ is not integral) while   time-reversal invariance is still maintained \cite{maciejkoFTI,maciejko_model,maciejko2015,swingle2011,YW13a}. Such topological phases are called ``fractional topological insulators'' (FTI).  In Ref.~\onlinecite{maciejkoFTI}, FTIs were obtained via parton constructions where the partons themselves carry fractional EM electric charges, and the internal gauge fields are in the Coulomb phase where gauge fluctuations are weak and the photon(s) are gapless. In a different non-fractionalized state where $U(1)\times U(1)\rtimes Z_2$ symmetry is considered \cite{bti6},  the $\Theta$ term may signal a mutual Witten effect where a monopole of one U(1) gauge group induces an electric charge of another $U(1)$ gauge group.

   In the following, we will   explore   {FTI}s via the parton construction introduced in Sec.~\ref{sec:gauge_strc}.   However in Sec.~\ref{sec_FTI_Coulomb}, we shall first study  charge lattices for which all partons occupy topological insulator bands and the internal gauge fields are in the Coulomb phase. The resulting state is a non-fractional topological insulator (i.e., $\Theta=\pi$) and there are two massless gauge bosons in the bulk.  In order to obtain a gapped bulk and a fractional $\Theta$, in Sec.~\ref{sec_FTI} we again assume that all partons occupy topological insulator bands and  then we condense certain composites (see $\varphi_1$ and $\varphi_2$ of {FTI} in Table~\ref{table:em}). We focus on a concrete example and show that the  resulting state is a  {FTI} with time-reversal symmetry and $\Theta=\frac{\pi}{9}$ (c.f. Eq.~(\ref{eqn:new_Theta_reduced})). As a side result,  in Appendix \ref{appendix_Gamma_trivial_1}, we show that the ans\"atze in which all partons are in a topologically trivial band structure always gives a topologically trivial state with $\Theta=0$ regardless of the condensate structure.

      \begin{widetext}

  \subsection{Topological insulators in the Coulomb phase}\label{sec_FTI_Coulomb}

 In the following, we consider partons occupying non-trivial 3D topological insulator bands (i.e., $\theta=\pi$). Previously, it was  shown that partons with $\theta=\pi$ can potentially support {a} fractional $\Theta$ angle if {the} Coulomb phase is considered, and a special parton representation of an electron is used~\cite{maciejkoFTI}.     
 In the Coulomb phase, the dynamical gauge fields $a_\mu$ and $b_\mu$ are weakly fluctuating and non-compact; hence, the standard perturbative analysis is applicable. 
  Integrating {out} the partons {to quadratic order in the gauge fields~\cite{Qi2008}, we obtain the following effective action $S_{\rm eff}$:}
  \begin{align}
\!S_{\rm eff}\!=\!& \!\int\! d^4x\frac{\theta}{32\pi^2}\left(g_a f^a_{\mu\nu}\!+\!eG_{\mu\nu}\right) \! \!\left(g_af^a_{\lambda\rho}\!+\!eG_{\lambda\rho}\right)\!\ep 
+ \! \!\int \!d^4x\frac{\theta}{32\pi^2}\!\!\left(-g_af^a_{\mu\nu}\!-\!g_bf^b_{\mu\nu}+eG_{\mu\nu}\right)\nonumber\\
&\times \! \!
 \left(-g_af^a_{\lambda\rho}\!-\!g_bf^b_{\lambda\rho}\!+\!eG_{\lambda\rho}\right)\!\ep+\!\!\int d^4x\frac{\theta}{32\pi^2}\left(g_bf^b_{\mu\nu}-eG_{\mu\nu}\right) \! \!\left(g_bf^b_{\lambda\rho}-eG_{\lambda\rho}\right)\ep+S_\text{Maxwell}\,,
  \end{align}
  \end{widetext}
where  $\theta=\pi$. The quantities $f^a_{\mu\nu}=\partial_\mu a_\nu-\partial_\nu a_\mu$ and $f^b_{\mu\nu}=\partial_\mu b_\nu-\partial_\nu b_\mu$  are field strength tensors of $a_\mu$ and $b_\mu$ respectively.  Both $a_\mu$ and $b_\mu$ are smooth variables and do not support monopole configurations. $G_{\mu\nu}$ is defined as: 
$G_{\mu\nu}=F_{\mu\nu}-\frac{2\pi}{e}S_{\mu\nu}$, where $F_{\mu\nu}=\partial_\mu A_\nu-\partial_\nu A_\mu$, $A_\mu$ is smooth external EM field, and the tensor $S_{\mu\nu}$ forms  the EM monopole current via: 
$M_\mu=\frac{1}{2}\ep \partial_\nu S_{\lambda\rho}\,
$.  The constant $e^2$ denotes the fine structure constant of  the EM field $A_\mu$. The coupling constants  $g_{a,b}$ of the $a_\mu$ and $b_\mu$ gauge fields are written explicitly and $0<g_a,g_b\ll 1$ in the Coulomb phase.   $S_\text{Maxwell}$ includes all non-topological terms (Maxwell-type) of $a_\mu,b_\mu$ and $A_\mu$.   Since both $a_\mu$ and $b_\mu$ are smooth variables, all terms of the form  $f^a\wedge f^a$, $f^a\wedge f^b$, and $f^b\wedge f^b$ are   total-derivative terms that can be neglected in the bulk effective field theory. The term $-\frac{4\theta \,e \,g_b}{32\pi^2}f^b_{\mu\nu} G_{\lambda\rho}\ep=\frac{\theta g_b}{\pi}M_\mu b_\mu$  implies that $M_\mu$ carries integer gauge charge of the $U(1)_b$ gauge group by noting that $\frac{\theta}{\pi}=1$. 
As such, after integrating $a_\mu,b_\mu$ $S_{\rm eff}$ reduces to:
\begin{align}
S_{\rm eff}=&\frac{\Theta e^2}{32\pi^2}\int d^4x G_{\mu\nu}G_{\lambda\rho}\ep+\cdots\,,\label{Effective_Action_Coulomb}
\end{align}
where $\Theta=3\pi$. The terms represented by $\cdots$ include the long-range Coulomb interactions between the monopole currents $M_\mu$ mediated by the $b_\mu$-photons, and other non-topological terms.  Since the periodicity of $\Theta$ is still $2\pi$ in the absence of charge fractionalization, $\Theta$ reduces to  $\pi$ by a $2\pi$ periodic shift.    In summary, the resulting state shows a $\Theta$ angle that is the same as a free-fermion topological insulator. The bulk  admits   two gapless, electrically neutral excitations, i.e., photons of the $U(1)_a$ and $U(1)_b$ gauge fields.

  \subsection{Fractional topological insulators in the composite condensation phase}\label{sec_FTI}

The charge  {lattice} in Sec.~\ref{sec_FTI_Coulomb} was obtained from the assumptions that (i)  {partons} occupy $\theta=\pi$ topological insulator bands, and (ii) the internal gauge fields are in the Coulomb phase. However, the  {resulting} phase supports a non-fractional $\Theta=\pi$ angle and the bulk spectrum is  gapless. In the following we consider composite condensation phases as discussed in Sec.~\ref{sec:gauge_strc} instead of the Coulomb phase. When the partons are in topological insulator bands the {resulting} phase can support fractionalized $\Theta$ angles and a \emph{fully} gapped bulk.  

 {Let} us start with the scenario that all partons occupy topological insulator bands (i.e., $\theta=\pi$) and then consider composite condensations. One can prove that parameters $u,v,u',v',q,q'$  must be even:
\begin{align}
u,v,u',v',q,q'\in\Z_{\rm even}\label{uvuvqq_even}\,{,}
\end{align}
in order to satisfy the set of constraints given by Eqs.~(\ref{constr_1},\ref{constr_3}).   {We obtain the following relations} via Eqs.~(\ref{eqn:transition_matrix},\ref{eqn:transition_matrix1},\ref{Nnf1}) ($\theta=\pi$):
\begin{align}
n^{f1}-n^{f2}=&N_a-N^a_m-\frac{1}{2}N^b_m\,,\label{N_a_n1n2}\\
n^{f3}-n^{f2}=&N_b-N^b_m-\frac{1}{2}N^a_m+M\,,\label{N_b_n1n2}\\
n^{f1}+n^{f2}-n^{f3}=&N_A+N^b_m-\frac{3}{2}M\,.\label{NA_is_integer}
\end{align}
In order to see whether or not there is charge fractionalization  (Definition~\ref{crt_charge_1}), we may check the value of $Q$ defined in Eq.~(\ref{NE1}) when $M=0$. Then, Eq.~(\ref{NA_is_integer}) indicates that $N_A$ is always integer-valued when $M=0$ by noting that $n^{fi}$ and $N_m^b$ are integer-valued.  

 Thus, we should further check whether or not  $Q_{\rm Debye}$ defined in Eq.\eqref{NE_screen} is fractional when $M=0$.  In principle, one may deduce $\Theta$ as a function of the parameters $(u,v,u',v',q,q',\theta)$. However, such a generic discussion is technically intricate and not illuminating.   Instead, we will proceed further with a concrete example as a proof of principle (see Table~\ref{table:em}): $
u=2,v=2,u'=4,v'=10,q=2,q'=2\,$.  
In terms of the matrix notation defined in Eq.~(\ref{define_KNq1}), we have:
\begin{align}
K=\bpm 2 & 2 \\ 4& 10\epm,  ~~~~\mathbf{q}=\bpm 2\\2\epm\,.\label{define_KNq12345}
\end{align}
From Table~\ref{table:em}, we see that $\varphi_1$ is a bosonic bound state of two $a_\mu$ magnetic monopoles,  two $b_\mu$ magnetic monopoles, one $f^1$ parton,  four $f^2$ partons, and  one $f^3$ parton. $\varphi_2$ is a bosonic bound state of four $a_\mu$ magnetic monopoles, ten $b_\mu$ magnetic monopoles,   nine $f^2$ partons, and three hole-like $f^3$ partons.

 By using Eq.~(\ref{NabM}), it is straightforward to work out the relation between $N_{a,b}$ and $M$:
\begin{align}
N_a=\frac{4}{3}M\,,~~~~~ N_b=-\frac{1}{3}M\,{,}\label{example_NaNb}
\end{align}
which must be satisfied for all excitations. Plugging Eq.~(\ref{example_NaNb}) into Eqs.~(\ref{N_a_n1n2},\ref{N_b_n1n2}), we end up with 
\begin{align}
n^{f1}-n^{f2}=&\frac{4}{3}M-N^a_m-\frac{1}{2}N^b_m\,,\label{N_a_n1n2111}\\
n^{f3}-n^{f2}=&\frac{2}{3}M-N^b_m-\frac{1}{2}N^a_m\,.\label{N_b_n1n2111}
\end{align}
It is obvious that the single parton $f^i$ whose charges are shown in Table~\ref{table:em} is confined since Eqs.~(\ref{N_a_n1n2111},\ref{N_b_n1n2111}) are not satisfied simultaneously. 

\begin{figure}[t]
\centering
\includegraphics[width=8.6cm]{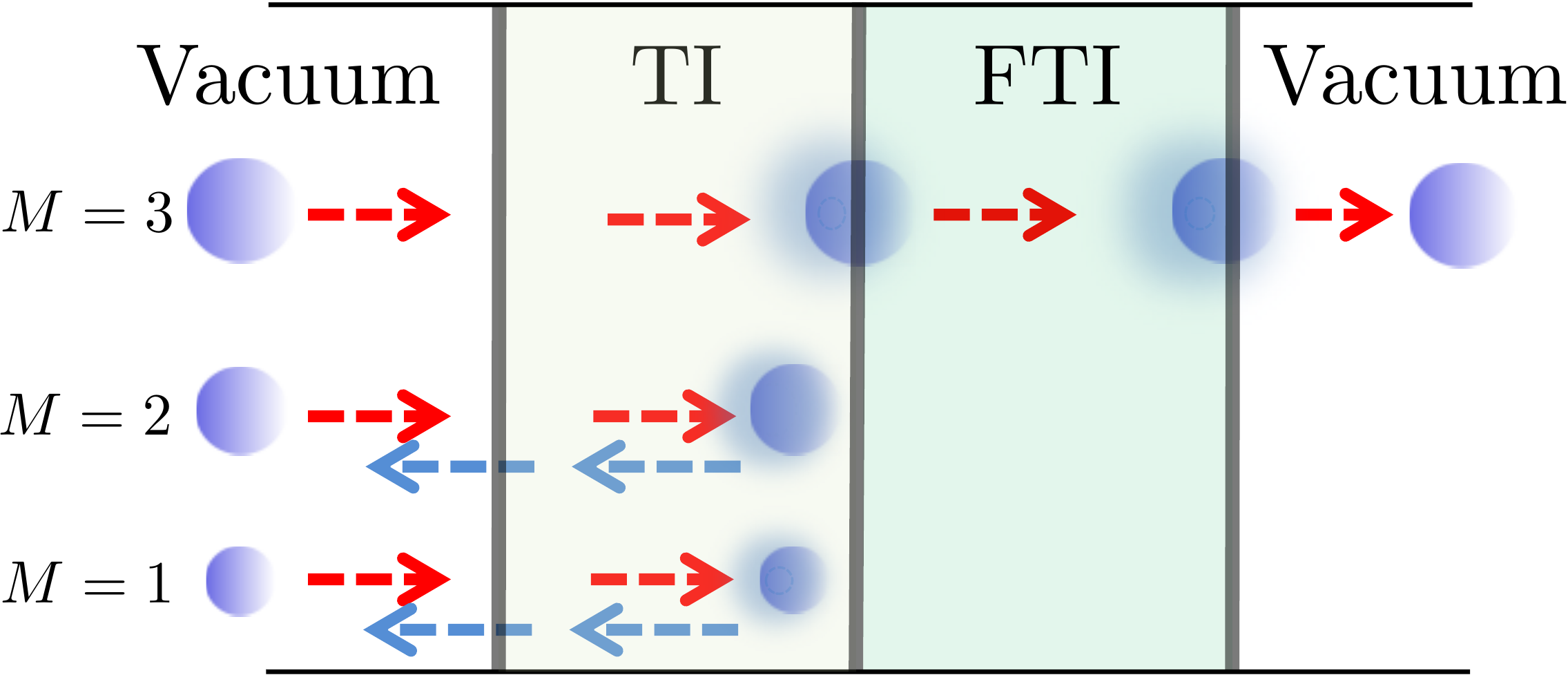}
\caption{(Color online) Throwing three external EM magnetic monopoles  (denoted by blue balls) in vacuum into topological materials TI and {FTI}. Only {the} EM magnetic monopole with {$3k$, $k\in\Z$} magnetic charge shown in (\ref{double_m_charge}) can penetrate the {FTI} boundary in our example. An EM magnetic monopole with $M=1,2$ will be completely reflected on the {FTI} boundary, as shown by the leftwards arrows. The shadow of each ball pictorially denotes the polarization charge cloud induced by the Witten effect.}
\label{figure_ti}
\end{figure}

By noting that   $n^{fi}, N_m^a,N_m^b,M$ are integer-valued, Eqs.~(\ref{N_a_n1n2111},\ref{N_b_n1n2111}) require that:
\begin{align}
&N_m^a, N_m^b\in\Z_{\rm even}   \,;~~~\,M=0,\pm3,\pm6,\pm9,\cdots\,.\label{double_m_charge}
\end{align}
Therefore, the quantization of $M$ is modified compared to the usual quantization $M=0,\pm1,\pm2{,\ldots}$ found in the vacuum and a non-fractionalized TI.  A direct consequence is that an $M=1$ particle is not allowed to pass through {the} {FTI}, which is   illustrated in Fig.~\ref{figure_ti}. 

We may use $n^{f1}$, $n^{f2}$, $n^{f3}$, and $M$ to uniquely label all excitations. Solving Eqs.~(\ref{N_a_n1n2111},\ref{N_b_n1n2111}) gives rise to: 
\begin{align}
&N_m^a=\frac{4}{3}M-\frac{4}{3}n^{f1}+\frac{2}{3}n^{f2}+\frac{2}{3}n^{f3}\,,\label{equation_NMA}\\
&N_m^b=\frac{2}{3}n^{f1}+\frac{2}{3}n^{f2}-\frac{4}{3}n^{f3}\,.\label{equation_NMB}
\end{align}
By using the above two equations, $N_A$ in Eq.~(\ref{NA_is_integer}) and $Q_{\rm Debye}$ in Eq.~(\ref{NE_screen})  can be expressed as: 
\begin{align}
&N_A=\frac{1}{3}n^{f1}+\frac{1}{3}n^{f2}+\frac{1}{3}n^{f3}+\frac{3}{2}M\,,\\
&Q_{\rm Debye}=-2n^{f1}+\frac{2}{3}n^{f2}+\frac{4}{3}n^{f3}+\frac{16}{9}M\,.
\end{align}
The net EM electric charge $Q$ is defined as $N_A-Q_{\rm Debye}$ and thus  is given by: 
\begin{align}
Q=\frac{7}{3}n^{f1}-\frac{1}{3}n^{f2}-n^{f3}-\frac{5}{18}M\,.\label{Q_expression}
\end{align}
Thus, the quantization of $Q$ is given by: (see Appendix \ref{appendix_gamma_stat4321_new})
\begin{align}
&Q=0,\pm\frac{1}{3},\pm\frac{2}{3},\pm1\,,\cdots \,\,\,\text{ when } \frac{M}{3}=0,\pm2,\cdots\,,\label{Q_fractional_M0}\\
&Q=0,\pm\frac{1}{6},\pm\frac{3}{6},\pm\frac{5}{6}\,,\cdots \,\,\, \text{ when } \frac{M}{3}=\pm1,\pm3,\cdots.\label{Q_fractional_M1}
\end{align}
Eq.~(\ref{Q_fractional_M0}) indicates that the intrinsic excitations of {the} {FTI} (Definition~\ref{dfn_intrinsic_exc}) carry $1/3$ quantized EM electric charge. In other words,  the {FTI} bulk supports charge fractionalization (Criterion \ref{crt_charge_1}). Due to the quantization of $M$ in Eq.~(\ref{double_m_charge}), Criterion \ref{crt_charge_2} is automatically satisfied.  

Eq.~(\ref{Q_fractional_M1}) indicates that the 2D $(M,Q)$ lattice  is tilted by  an angle $\frac{5}{18}M$. More precisely, an axion angle $\Theta$ can be defined as: $\Theta=-\frac{5}{9}\pi$ 
by identifying $- \frac{5}{18}M=\frac{\Theta}{2\pi}M$.  This $M$-dependent EM electric charge is a known  consequence of the Witten effect \cite{witten1,Qi2008,franz}.

{The} self-statistics of excitations (i.e., either fermionic or bosonic) can also be derived as a function of $(n^{f1},n^{f2},n^{f3},M)$. For this purpose, let us start with $\Gamma$ defined in Eq.~(\ref{eq:stat123}) and take Eq.~(\ref{double_m_charge}) into account. Therefore, the first two terms of Eq.~(\ref{eq:stat123}) are even and can be removed giving:
 \begin{align}
\Gamma=&(M+1)(n^{f1}+n^{f2}+n^{f3})\,,\label{eq:stat4321}
\end{align}
where $-n^{f3}$ is also changed to $n^{f3}$ leaving the even-odd property of $\Gamma$ unaltered.
 In analogy to a TI, time-reversal symmetry should also be {maintained}. From the point of view  {of the charge lattice}, time-reversal symmetry is a reflection  {symmetry} $M\rightarrow -M$  {that keeps the} net EM electric charge and self-statistics invariant{:} $Q\rightarrow Q${,} $\Gamma\rightarrow \Gamma+\text{even integer}$.   {One possible definition of time-reversal symmetry that satisfies these properties is as follows,}
 \begin{align}
& \mathcal{T}n^{f1} \mathcal{T}^{-1}= n^{f1}-\frac{1}{3}M\,,~ \mathcal{T}n^{f2} \mathcal{T}^{-1} =n^{f2}-\frac{2}{3}M\,, \\
 & \mathcal{T}n^{f3} \mathcal{T}^{-1}= n^{f3}\,, ~~\mathcal{T}M \mathcal{T}^{-1}=-M\,,
 \end{align} 
where $\mathcal{T}$ denotes  {the} time-reversal operator. It can be  {verified} that $Q$ is invariant and $\Gamma$ is only changed by  {an} even integer, thus leaving its even-odd property unaltered.  {Using} the above transformations, we may also derive the transformations below: 
\begin{align}
 \mathcal{T}N^a_m \mathcal{T}^{-1}=N_m^a-\frac{8}{3}M\,,~ \mathcal{T}N^b_m \mathcal{T}^{-1}=N_m^b-\frac{2}{3}M\,.
 \end{align} 
The shifted amounts $-\frac{8}{3}M$ and $-\frac{2}{3}M$ are even integers, which guarantees the transformed $N^{a,b}_m$ are still even as required by Eq.~(\ref{double_m_charge}).  
 {A time-reversed excitation is still an excitation, in the sense that the} transformed electric and magnetic charges also satisfy all equations that are satisfied by the excitation before  {time reversal}. Geometrically,  {this} means that after the above transformations, the new particle is still on the 4D charge lattice. Furthermore, $Q$ and  {$\Gamma$} are unchanged.  From this geometric point of view, the time-reversal symmetry defined above effectively acts like a subgroup of  {the point group} of the 4D charge lattice. 
 \begin{table}[tbp]
\centering
\caption{Examples of  excitations in our {FTI}.  The electric and magnetic charges are explicitly shown. ``F'' is short for ``fermionic'' where $\Gamma$ is odd.   ``B'' is short for ``bosonic'' where $\Gamma$ is even. We call an elementary charge an intrinsic excitation (Definition~\ref{dfn_intrinsic_exc}) that {carries} $Q=1/3$ or $Q=2/3$ EM electric charge, in analogy to the fractionalized charge excitation{s} in  {the} $\nu=1/3$ {FQH} state. {The} two elementary charges in the Table are just two concrete examples, and there are many other excitations that carry $Q=1/2,2/3$ and $M=0$. {The} elementary EM monopole is an excitation that carries the minimal nonzero EM magnetic charge $M=3$ and do{es} not contain any partons (i.e., $n^{fi}=0$, $\forall i=1,2,3$).  {A} nonzero $M$ can be externally added into the bulk in order to probe the EM response (see Definition~\ref{dfn_intrinsic_exc}). {The} minimal quantum of charge fractionalization  is $\frac{1}{3}$ determined by the {intrinsic excitations}, i.e., Eq.~(\ref{Q_fractional_M0}) rather than Eq.~(\ref{Q_fractional_M1}).}
\label{table:FTI}
\begin{tabular}{c|cccccccccc}\hline

\hline

\hline
\hline
 & $n^{f1}$ & $n^{f2}$ & $n^{f3}$ & $M$ & $N_A$ & $Q_{\rm Debye}$ & $Q$ & $N_m^a$ & $N_m^b$ & $\Gamma$ \\\hline  
\begin{minipage}[t]{0.77in}Elementary charge \end{minipage}
  & $1$ & $0$ & $2$ & $0$ & $1$ & $\frac{2}{3}$ & $\frac{1}{3}$ & $0$ & $-2$ & F\\\hline  
\begin{minipage}[t]{0.77in}Elementary charge\end{minipage}
  & $2$ & $9$ & $1$ & $0$ & $4$ & $\frac{10}{3}$ & $\frac{2}{3}$ & $4$ & $6$ & B\\\hline  
Electron & $1$ & $1$ & $1$ & $0$ & $1$ & $0$ & $1$ & $0$ & $0$ & F \\\hline
 \begin{minipage}[t]{0.78in}Elementary EM monopole\end{minipage}  &  $0$& $0$ & $0$ & $3$ &$\frac{9}{2}$  &$\frac{16}{3}$  &$-\frac{5}{6}$  &$4$  &$0$  & B\\\hline
 \begin{minipage}[t]{0.77in}An example with $M=3$ \end{minipage} &   $1$&  $1$ &  $1$ & $3$  &  $\frac{11}{2}$ &  $\frac{16}{3}$ &   $\frac{1}{6}$&  $4$ &   $0$&B\\\hline
 \begin{minipage}[t]{0.77in}An example with $M=3$ \end{minipage} &   $2$&  $1$ &  $0$ & $3$  &  $\frac{11}{2}$ &  $2$ &   $\frac{7}{2}$&  $2$ &   $2$&B\\\hline
 \begin{minipage}[t]{0.77in}An example with $M=6$ \end{minipage} &   $1$&  $1$ &  $1$ & $6$  &  $10$ &  $\frac{32}{3}$ &   $-\frac{2}{3}$&  $8$ &   $0$&F\\\hline
 \begin{minipage}[t]{0.77in}An example with $M=6$ \end{minipage} &   $2$&  $1$ &  $0$ & $6$  &  $10$ &  $\frac{22}{3}$ &   $\frac{8}{3}$&  $6$ &   $2$&F\\ \hline\hline

\hline

\hline
\end{tabular}
\end{table}

\begin{figure}[t]
\centering
\includegraphics[width=8.5cm]{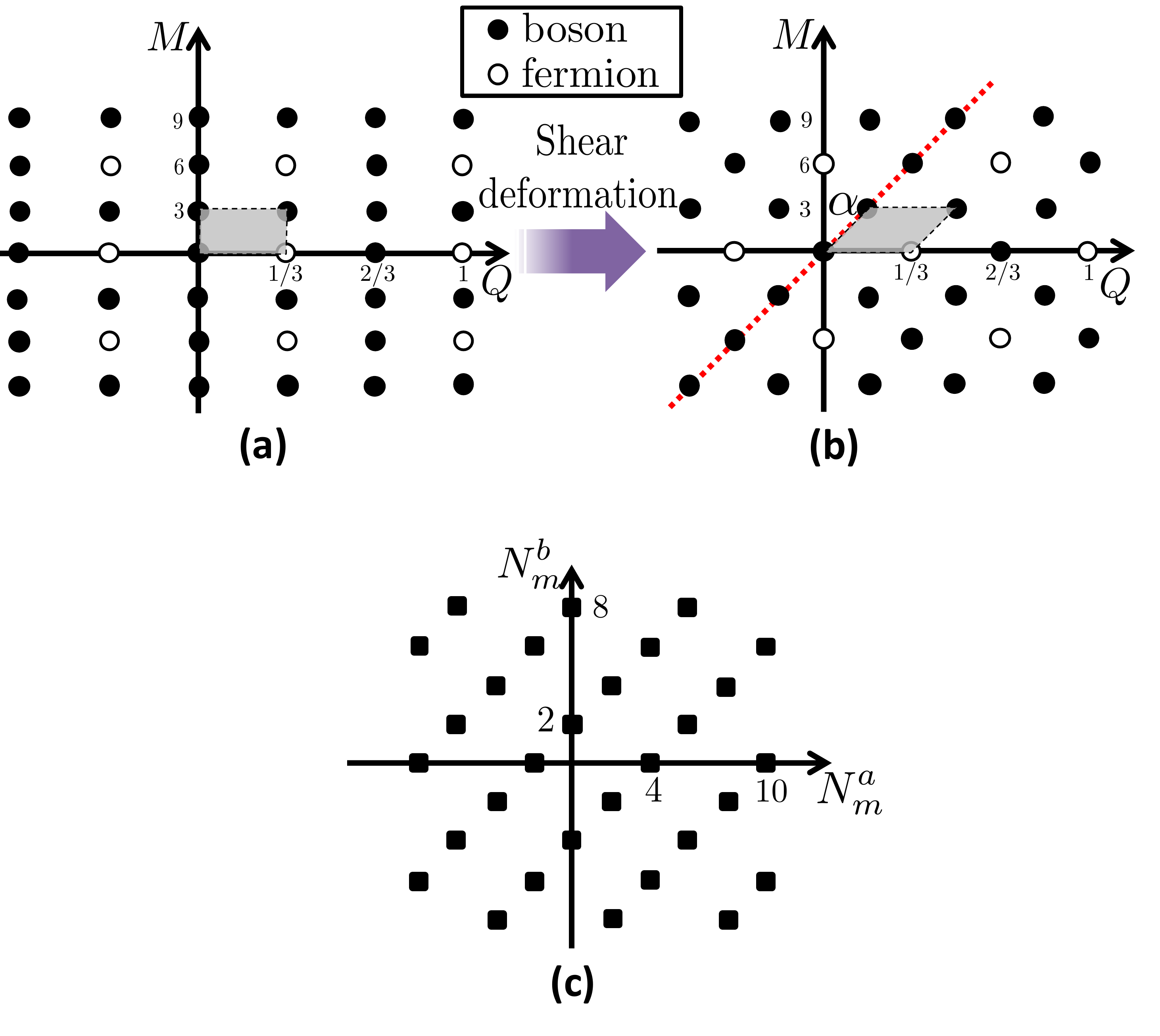}
\caption{(Color online)
Self-statistics distribution on the 4D charge lattice. (a) Self-statistics distribution as a function of $M$ and $Q$ by turning off $\Theta$  in Eq.~(\ref{eq:stat4321_new}).  (b) Self-statistics distribution as a function of $M$ and $Q$ as shown  in Eq.~(\ref{eq:stat4321_new}).   The allowed values of $M$ and $Q$ are determined by Eqs.~(\ref{Q_fractional_M0},\ref{Q_fractional_M1}).     Geometrically, $\Theta=2\pi\tan\alpha$, where $\tan\alpha=\frac{1/6}{3}=1/18$. Thus, $\Theta$ angle can be viewed as a consequence of a shear deformation from (a) to (b). During the shear deformation, the area of ``Dirac unit cell'' (denoted by the shaded area) is invariant. Since the charge lattice is 4D (Definition \ref{dfn_excitation}), each site in  (b) on the $(M-Q)$ parameter space corresponds to more than one excitation.  An example is shown in (c), where $N^a_m,N_m^b$ are used to {label} excitations that have the same $Q$ and $M$: $Q=\frac{1}{6},M=3$.}
\label{figure_theta}
\end{figure}

   An important result is that the {FTI} with $\Theta=-\frac{5}{9}\pi$ is actually topologically equivalent to the {FTI} with $\Theta=\frac{1}{9}\pi$ by  {a} periodic shift. The minimal choice of $\Theta$ for our {FTI} phase is given by:
\begin{align}
\Theta=\frac{1}{9}\pi \text{  mod  }\frac{2}{9}\pi\,.\label{eqn:new_Theta_reduced}
\end{align}
 {To understand} this result,  let us revisit the self-statistics $\Gamma$ in Eq.~(\ref{eq:stat4321}). In fact, $\Gamma$ can be reformulated as a unique function of $M$ and $Q$: (see Appendix \ref{appendix_gamma_stat4321_new} for details)
\begin{align}
\Gamma=&3(M+1)(Q-\frac{\Theta}{2\pi}M)\,,\label{eq:stat4321_new}
\end{align}
where $\Theta$ is given by Eq.~(\ref{eqn:new_Theta_reduced}). We manifestly see that  the even-odd property of $\Gamma$ is unaltered by the \emph{minimal} shift $\Theta\rightarrow \Theta+\frac{2}{9}\pi$.   $\Gamma$ is pictorially illustrated in Fig.~\ref{figure_theta}(b) and we can see  that  geometrically, $\Theta$ describes how tilted the charge lattice is with respect to its initial orientation (Fig.~\ref{figure_theta}(a)). To illustrate, the red dashed line in Fig.~\ref{figure_theta}(b) can be either more or less tilted with respect to the vertical axis via a shear deformation. 
  The charge lattice  [Fig.~\ref{figure_theta}(b)] with a nonzero $\Theta$ can be obtained through such a shear deformation from the non-tilted charge lattice [Fig.~\ref{figure_theta}(a)]. Since $\Theta=\frac{1}{9}\pi$,  the charge lattice shown in Fig.~\ref{figure_theta}(b) is time-reversal invariant ($Q\rightarrow Q, M\rightarrow -M, \Gamma\rightarrow \Gamma+\text{even integer}$), which can be viewed as a reflection symmetry about $Q$-axis.

   It is obvious that the entire charge lattice [Fig.~\ref{figure_theta}(b)] as well as the self-statistics distribution is unaltered if we further increase $\Theta$ by  $\frac{2}{9}\pi$ (i.e. increase $\tan\alpha$ by $1/9$). For this reason,  $\Theta$ is well defined only mod  $\frac{2}{9}\pi$ as shown in Eq.~(\ref{eqn:new_Theta_reduced}). For example, the bosons on the site $(\frac{1}{6},3)$ are shifted to the bosons on the site $(\frac{1}{2},3)$; the fermions on the site $(0, 6)$ are  shifted to the fermions on the site $(\frac{2}{3},6)$.  Furthermore, since the charge lattice is actually 4-dimensional (Definition \ref{dfn_excitation}), each lattice site of Fig.~\ref{figure_theta}-(b) actually corresponds to many excitations that are different from each other by $N^a_m,N^b_m$ as shown in Fig.~\ref{figure_theta}-(c) where $Q=\frac{1}{6},M=3$ is illustrated. The lattice sites in  Fig.~\ref{figure_theta}-(c) follow a simple relation: $({N^b_m-N^a_m-1})/{3}\in\Z$ where $N^a_m,N^b_m$ are even. (see Appendix \ref{appendix_gamma_stat4321_new} for details.)

 Experimentally, one may understand the physics of $\Theta$ via the surface quantum Hall effect on a surface with broken time-reversal symmetry. For example, by placing a ferromagnetic thin film on top of the surface of {a} {FTI}, we may observe {a} Hall effect with Hall conductance  $\sigma_H=\frac{\Theta}{2\pi}\frac{e^2}{h}$ \cite{Qi2008}:
 \begin{align}
\sigma_H=\left(\frac{1}{18}+\frac{n}{9}\right)\frac{e^2}{h}\,,n\in\Z\,,
\end{align} 
where $h$ is the usual Planck constant.  It should be kept in mind that, although the minimal nonzero $\sigma_H$ is $\frac{1}{18}$, the corresponding charge induced by the Hall response is not $\frac{1}{18}$ since the quantization of $M$ in Eq.~(\ref{double_m_charge}) is modified from its non-fractionalized value.  More precisely, an external EM monopole with  $M$ magnetic charge can be viewed as $2\pi M$ EM magnetic fluxes threading the surface \cite{bti5}. By using {the} Laughlin argument, the surface will generate $\frac{1}{18}M$ response charge once the EM monopole penetrates the surface. Since the minimal nonzero $M$ is $3$ due to Eq.~(\ref{double_m_charge}), the minimal surface response charge is $\frac{1}{18}\times 3=\frac{1}{6}$ rather than $1/18$.  Further, the $\frac{1}{6}$ charge will be attached onto the EM magnetic monopole that moves into the {FTI} bulk, which renders the Witten effect \cite{witten1,Qi2008,franz}. This phenomenon is nontrivial in a sense that the $\frac{1}{6}$ charge cannot be formed by the bulk intrinsic excitations (Definition \ref{dfn_intrinsic_exc}) whose $Q$ is quantized at $\frac{1}{3}$ due to Eq.~(\ref{Q_fractional_M0}).

{An} {FTI} can be viewed as a symmetry-enriched topological phase  {(SET)} which {is} characterized by both bulk topological order (TO) data and a symmetry action. In our case, the latter is encoded by the structure of the charge lattice in Fig.~\ref{figure_theta}. The former is given by {the set of} all intrinsic excitations (Definition~\ref{dfn_intrinsic_exc}, i.e., all sites along the $Q$-axis in Fig.~\ref{figure_theta})  and also loop excitations (Definition~\ref{dfn_loops}). With these preliminaries, we may discuss  the consequence of stacking operations in the context of topological order \cite{string8,wen_stacking}.  {Stacking operations}, denoted as $\boxtimes$, form a monoid that does not  contain inverse elements. It is known that stacking two TIs leads to  {the topologically trivial vacuum state}:  $\text{TI}\boxtimes \text{TI}=\text{Vacuum}
$. Let us stack a 3D TI and a 3D {FTI} together. The  {resulting} phase is a TO: $\text{FTI}\boxtimes \text{TI}=\text{TO}
$.  In other words, the stacking operation removes the nontrivial Witten effect of the {FTI}, rendering a  {state with} pure topological order.  This can be understood  {in} two steps. First, 
since the bulk intrinsic excitations of a TI only contain electron excitations, the above stacking operation indeed does not  change the TO of the {FTI}. Second, in the stacked phase, the net EM electric charge $Q$ is given by $Q=\left(\frac{\pi}{18}M+\frac{n}{3}\right)+\left(\frac{\pi}{2}M+n'\right) 
$ 
{with $n,n'\in\Z$}, where the first term is given by  {the} Witten effect of  {the} {FTI} while the second term is given by  {the} Witten effect of  {the} TI. Since the stacked phase  {is} formed by putting  {the} {FTI} and TI in  {the} same  {spatial} 3D region, the quantization of $M$ in Eq.~(\ref{double_m_charge}) still holds in the stacked phase. As a result,  {the electric charge} $Q$  {in} the stacked phase is given by: $Q=\frac{5}{9}M+\frac{n}{3}+n'$, where the $M$-induced charge $\frac{5}{9}M$ is quantized  {to} $5/3$. This charge can be  {completely} screened by $\frac{n}{3}+n'$, e.g.{,} $n=-2\,,\,n'=-1$. Thus, the charge lattice of the stacked phase is not tilted, meaning that  $\Theta=0$.

  In summary, the stacking of a {FTI} and a TI leads to a  {phase with} pure topological order where  {the} Witten effect is absent.  
We may also consider stacking  {two} {FTI}s: $
\text{FTI}\boxtimes \text{FTI}=\text{TO}\boxtimes \text{TO}$, which means that the stacked phase is a  purely topologically ordered phase where the charge lattice is not tilted and the topological order is given by $\text{TO}\boxtimes \text{TO}$. Surely, this is just an example while {it} is possible that other examples of {FTI} may  {produce different phases when} stacked together with TI or  {with} themselves.  {S}tacking operations in SET phases  {generally} change TO to a new topological order denoted as ``$\widetilde{\text{TO}}$''. For example, stacking two {FTI}s here gives rise to  $\widetilde{\text{TO}}=\text{TO}\boxtimes \text{TO}$.  In order to see if the  {resulting} phase is a new SET or not, one should further consider symmetry-respecting condensations that change  $\widetilde{\text{TO}}$ back to TO.  {In} this way, we may make progress toward the classification of SETs. As it is beyond the scope of the present work, we will leave this issue to further studies.

The above calculation is based on concrete numerical inputs (\ref{define_KNq12345}). As  {mentioned previously}, one may in principle generically deduce $\Theta$ as a function of the parameters $(u,v,u',v',q,q',\theta)$ that fully determine the two permissible composite condensations and the  {entire} bulk spectrum. In 2D, we know that some FQH states can be unified into Jain's sequence \cite{composite_Jain_1,composite_Jain_2} such that they can be understood in the composite-fermion theory with different microscopic designs of the composite particles. Our 3D composite particle theory is   similar to this 2D scenario: the $\Theta$ angle, and other properties of  composite condensation phases, are also determined by the different designs of composite condensations. Therefore, all phases constructed from composite condensations can be thought to form a sequence. We expect more studies in the future along this line of thinking will be helpful in uncovering  the physics of 3D Abelian topological phases of strongly interacting fermion systems.
   
 \subsection{Deconfined descrete gauge subgroup $\Z_2\times \Z_6$: Abelian topological order in the bulk}  \label{sec_FTI_plus}

   The FTI state obtained in Sec.~\ref{sec_FTI} supports fractionalized intrinsic excitations as indicated by Eq.~(\ref{Q_fractional_M0}) and the texts around it. Ref.~\onlinecite{swingle2011} ever pointed out that FTIs necessarily has a fractionalized bulk. Therefore, our construction is consistent to the claim. Usually, a fractionalized gapped bulk can be understood as the presence of a topological order of some form. To see more clearly the exact form of the topological order of our FTI, let us start with the $K$ matrix in Eq.~(\ref{define_KNq12345}). By using two independent unimodular matrices (i.e., $\Omega$ and  $W$ that will be discussed in details in Sec.~\ref{sec:charlos_defect_symmetry}), we may diagonalize $K$:
\begin{align}   
& \Omega K W^T=\bpm 2 & 0 \\ 0& 6\epm, \\
 &\text{where }\Omega=\bpm 1&0\\-2&1  \epm\,,\,W=\bpm 1&0\\-1&1  \epm\,.
\end{align}
In the new basis, it is clear that the bosonic sectors of the ground state are described by deconfined  $\Z_2\times\Z_6$ gauge group. In other words, the maximal torus $U(1)\times U(1)$  of the  $SU(3)$ gauge group of the parton construction is confined \emph{except} the $\Z_2\times\Z_6$ gauge subgroup. In Ref.~\onlinecite{swingle2011}, the discrete gauge group $\Z_2$ arises since the choice of parton mean-field Hamiltonian explicitly breaks the original pseudospin $SU(2)$ gauge group down to $\Z_2$ subgroup. However, in our FTI state, the discrete gauge subgroup arises from the the deconfined subgroup of a confined non-Abelian gauge group, physically due to the condensation of composites that contain magnetic monopoles.

   
 \section{Charge-loop excitation symmetry and its relation to extrinsic twist defects}\label{sec:charlos_defect_symmetry}
 In Sec.~\ref{sec:theta}, we have explored the  axion angle  of the charge lattice with composite condensation. In this section, we will explore the charge-loop excitation symmetry based on the composite particle theory introduced in Sec.~\ref{sec:gauge_strc}.
 
 The topological $BF$ field theory (\ref{BFaction}), which is derived from the two permissible composite condensates, only captures the statistical interaction between particles that carry $N^a_m,N^b_m$ magnetic charges and loops that carry $\Phi_e^a,\Phi^b_e$ electric fluxes. Specifically, several important   properties of composites, such as the self-statistics  $\Gamma$ in Eq.~(\ref{eq:quantum_stat}), and the net EM electric charge $Q$ in  Eq.~(\ref{NE1}),   are not encoded in Eq.~(\ref{BFaction}). However, the topological $BF$ field theory reproduces $Q_{\rm Debye}$, an important part of $Q$.   
In this section, we further study the topological $BF$ field theory and show that it serves as a useful platform to study ``\emph{Char}ge-\emph{Lo}op \emph{E}xcitation \emph{S}ymmetry'' (abbreviated as ``$\mathsf{Charles}$'', see Definition~\ref{dfn_charlos}) that can be viewed as a 3D generalization of  ``anyonic symmetry'' \cite{Teo2015} (or ``topological symmetry'' in Ref.~\onlinecite{Barkeshli2014} and references therein) in 2D Abelian topological phases. We expect that 3D Abelian topological phases where loop excitations are allowed may host even more exotic physics if extrinsic twist defects are imposed, and anticipate that 3D charge-loop excitation symmetry will be a useful tool in future studies of such extrinsic defects.
 
  \subsection{Definition of $\mathsf{Charles}$}\label{sec:charlos}

 In 2D topological phases,  each point-like extrinsic twist defect is associated with an element of an anyonic symmetry group $G$.   The anyonic symmetry group is a finite group that acts to permute a subset of anyons in the parent TO phase while preserving all of the topological properties (topological spin, statistics) of the anyons (and sometimes their symmetry properties as well, e.g., their EM charge).   For example, the permutation of $e$ and $m$ particles in the 2D Wen-plaquette model (with $\Z_2$ TO)\cite{Wen2003} is a typical anyonic symmetry transformation. For this particular model this transformation can be realized by extrinsically imposing a lattice dislocation which enacts the  permutation of $e$ and $m$   when an anyon passes through a 1D  branch cut that terminates at the extrinsic point defect \cite{Kitaev2006,Bombin2010,You2012,Teo2015}. Interestingly, this quasiparticle permutation mechanism endows the dislocation with an attached non-Abelian object at the defect core, which opens up a possible new platform for topological quantum computation. Mathematically speaking, the incorporation of extrinsic defects into 2D Abelian topological phases described by a category theory $\mathcal{C}$, promotes  $\mathcal{C}$ to  a  $G$-crossed tensor category theory  $\mathcal{C}^{\times}_{G}$ \cite{Barkeshli2014,Teo2015}.

Let us briefly recall some properties of anyonic symmetry in 2D Abelian topological phases. As mentioned, these phases are described using Abelian Chern-Simons theory using the data in a symmetric, integer $K$-matrix. There is an important class of unimodular, integer transformations $W$ satisfying $WKW^{T}=K$ that act as the automorphisms of $K$ (or the automorphisms of the integer lattice, and dual/quasiparticle lattice, determined by $K$). These transformations relabel the different anyonic excitations, but most of them preserve the anyon type, and just attach local quasiparticles (e.g., attaching extra electrons). These trivial transformations are called the inner automorphisms ${\mathsf{Inner}(K)}$ and they form a normal subgroup of the full set of automorphisms $\mathsf{Auto}(K).$ The non-trivial anyonic relabeling symmetries are hence given by the group $G\equiv\mathsf{Outer}(K)=\frac{\mathsf{Auto}(K)}{\mathsf{Inner}(K)}.$ This captures the conventional anyonic symmetries that act as point-group operations on the quasiparticle lattice, although it leaves out possible non-symmorphic lattice operations or symmetries of stably-equivalent $K$-matrices \cite{Teo2015,Teo2014,cano2013}. We will not consider these more complicated possibilities for anyonic symmetries any further and leave their 3D generalization to future work. 

In order to generalize this discussion of anyonic symmetry and extrinsic defects to 3D, let us revisit some basic facts of excitations in our 3D fermionic gapped phase formed by two permissible composite condensates.  The 2D vectors $\mathbf{L}$ form a 2D loop-lattice in Definition~\ref{dfn_loops}. The 2D vectors $\mathbf{N}_m$ form a 2D  lattice which is a sublattice of the 4D charge lattice in Definition~\ref{dfn_excitation}.   As a whole, we may define a 6D \emph{charge-loop-lattice} (N.B., this is not the same 6D lattice mentioned earlier).  
\begin{dfn}
[{Charge-loop-lattice}] The charge-loop-lattice is a 6D lattice whose sites are given by the 6D lattice vector $\vec{V}=(N_A,\mathbf{N}^T_m,M,\mathbf{L}^T)=(N_A,N^a_m,N^b_m,M,\ell,\ell')$. Each site corresponds to a \emph{charge-loop composite}.\label{dfn_cloop}
\end{dfn}  In order to avoid confusions in terminology, the word ``composite,'' if used by itself, always denotes a point-like particle, unless otherwise specified.
The symmetry group  $\mathsf{Charles}$ is then defined as below:

\begin{dfn}
[{Charge-Loop Excitation Symmetry} ($\mathsf{Charles}$)] The charge-loop excitation symmetry group is a subset of the point group of the 6D charge-loop-lattice and corresponds to the following quotient group:
\begin{align}
\mathsf{Charles}=\frac{\mathsf{Auto}(K)}{\mathsf{Inner}(K)}\,,\label{auto_inner}
\end{align}
where $\mathsf{Auto}(K)$ is  the group  of generalized automorphisms of $K$.   $\mathsf{Inner}(K)$ is the group of generalized inner automorphisms of $K$, which is a subgroup of $\mathsf{Auto}(K)$.  Group elements of $\mathsf{Auto}(K)$ have the matrix representation  $\mathscr{G}=W\oplus \Omega$, where the two independent rank-two unimodular   matrices $W$ and $\Omega$ satisfy the following two conditions: 
\begin{align}
&(i).~~~~  \Omega K W^T=K\,,\label{eqn:auto}\\
&(ii).~~~~\Gamma(\cdots, {\mathbf{N}_m},\cdots)=\Gamma (\cdots,W^{-1}\mathbf{N}_m,\cdots)\,.\label{eqn:auto1}
\end{align}
Here, $\Gamma$ is the self-statistics of composites, which   is a function of lattice sites labeled by the 4D coordinates ($N_A,\mathbf{N}^T_m,M$).   In addition to conditions (i) and (ii), the group elements in $\mathsf{Inner}(K)$ have the property that    $W^{-1}\mathbf{N}_m-\mathbf{N}_m=K^T\,(n_1,n_2)^T$  and $\Omega^{-1}\mathbf{L}-\mathbf{L}=K(n_3,n_4)^T$, where $n_1,\cdots,n_4$ are integers. $n_1$ and $n_2$ are functions of  $\mathbf{N}_m,W$; $n_3$ and $n_4$ are functions of  $\mathbf{L},\Omega$. \label{dfn_charlos}
\end{dfn}
  Just like the 2D anyonic symmetry group,  the definition of $\mathsf{Charles}$ also involves the definitions of $\mathsf{Auto}(K)$ and $\mathsf{Inner}(K)$. One can prove that $\mathsf{Auto}(K)$ satisfies the usual  group axioms (identity element, inverse element, closure, associativity) and  that  $\mathsf{Inner}(K)$ is a normal subgroup of $\mathsf{Auto}(K)$, such that $\mathsf{Charles}$ forms a group.  Details of this proof can be found  in  Appendix \ref{appendix_proof_group}.

Physically,   group elements $\mathscr{G}=W\oplus\Omega$ in $\mathsf{Charles}$ correspond to  point group transformations: $({\mathbf{N}_m})_{\text{new}} = W^{-1} \mathbf{N}_m, \,({\mathbf{L}})_{\text{new}} = \Omega^{-1}\mathbf{L}$. 
  Conditions (i) and (ii) guarantee that the transformed charge-loop-lattice is   identical to the original one, which means that     $\mathsf{Charles}$ keeps not only the lattice geometry invariant, but also leaves all topological properties  of particle excitations and loop excitations (denoted by lattice sites) unaffected. Those topological properties include the self-statistics of particle excitations $\Gamma$, the charge-loop braiding statistics $\vartheta^{cl}$, and the Debye screening $Q_{\rm Debye}$.  However, there is a redundancy corresponding to  $\mathsf{Inner}(K)$ that should be removed.  $\mathsf{Inner}(K)$ includes all trivial transformations whose point-group effects are equivalent to effectively shifting  both $\mathbf{N}_m$ and $\mathbf{L}$ by   undetectable amounts (i.e., $\vartheta^{cl}=0 \text{ mod } 2\pi$ in Eq.~(\ref{eq:cltheta}); see also Sec.~\ref{sec:loop_loop}), and thereby must be modded out  from $\mathsf{Auto}(K)$ if we only want to keep non-trivial transformations. Again, the transformations in $\mathsf{Inner}(K)$ can be interpreted as changing the excitations by a trivial, topologically-undetectable charge or flux.

  In contrast to the 2D definition of ``anyonic symmetry'' where condition (i) (where the simpler structure only allows for $W=\Omega$) is enough to guarantee the invariance of the self-statistics of anyons,   one now needs condition (ii) in order to guarantee that the   self-statistics  of excitations on the 4D charge lattice remains invariant under $\mathsf{Charles}$ transformations.  The main reason for this is that  the self-statistics of an excitation (Definition~\ref{dfn_excitation}) cannot be captured by the topological $BF$ field theory. \emph{Whether or not $W$ satisfies   condition (ii) relies  on the specifics  of the parton decomposition, and  in the following subsections, we will assume condition (ii) is   satisfied.}
 
    \subsection{General theory of $\mathsf{Charles}$ and its tensor-network-type representation}\label{sec:general_charlos}
    It should be noted that $W$, $\Omega$, and $K$ in Definition~\ref{dfn_charlos} can be naturally generalized to arbitrary rank if a physical realization using  a scenario having any number of permissible composite condensates in Sec.~\ref{sec:bec} can be achieved. For example, one can consider a single composite condensate or three linearly independent condensates with entirely different parton constructions, which leads to a number $K\in\Z$ or a rank-three $K$ matrix respectively.  
    
     Before proceeding further, we introduce a simplified
notation that will be useful for subsequent discussions. The notation in the two-component $BF$ action (\ref{BFaction}), such as $\tilde{a}$ and $\tilde{b}$, comes from the specific physical realization described in Sec.~\ref{sec:bec}.   It is however    inconvenient for the purpose of generalizing $\mathsf{Charles}$. Thus,  in the current section (Sec.~\ref{sec:general_charlos}), we temporarily  use a new notation for the gauge fields: $b=(b^1,b^2,\cdots)$ and $a=(a^1,a^2,\cdots)$ where $b$ is a set of $2$-form Kalb-Ramond $U(1)$ gauge fields while $a$ is a set of $1$-form $U(1)$ gauge fields.   As a result, the topological $BF$ term is expressed as: 
 \begin{align}
 \frac{iK^{IJ}}{2\pi}\int b^{I}\wedge da^J=\frac{i}{2\pi}\int b^T\wedge K da
 \end{align} with a square matrix $K$ of rank $N$. The excitation terms in Eq.~(\ref{BFex}) are rewritten as: 
 \begin{align}
 \mathcal{S}_{ex}=i \int \mathbf{t}^T a\wedge \star  j+i \int\mathbf{L}^T  b \wedge \star  \Sigma\,,
 \end{align}
  where $\mathbf{t}=(t_1,t_2,\cdots)$ is an integer vector replacing the notation $\mathbf{N}_m$. Then, the charge-loop-lattice is formed by an $N$-dimensional charge lattice labeled by vectors $\mathbf{t}$ and an $N$-dimensional  loop-lattice labeled by vectors $\mathbf{L}$. Group elements of $\mathsf{Charles}$ are still denoted as ``$\mathscr{G}=W\oplus \Omega$'' with the transformations: $(\mathbf{t})_{\text{new}} = W^{-1}\mathbf{t}$ and  $(\mathbf{L})_{\text{new}} = \Omega^{-1}\mathbf{L}$.

  Let us consider some examples.  In Table~\ref{table:charlos}, all possible $\mathsf{Charles}$ groups are listed for a $1\times 1$ matrix $K\in\Z$.  From the table, we see that $\mathbb{Z}_2$ gauge theory in (3+1)D $(K=2)$ only has trivial $\mathsf{Charles}$, which is surprisingly different from  a deconfined $\mathbb{Z}_2$ gauge theory in (2+1)D (e.g., as appears in the Wen-plaquette model), where the $e\leftrightarrow m$ exchange process is an anyonic symmetry transformation. Nontrivial $\mathsf{Charles}$ arises   for $\mathbb{Z}_K$ gauge theory in (3+1)D \emph{only} when $|K|\geq 3$. For example, for $\mathbb{Z}_3$ gauge theory, the nontrivial element of $\mathsf{Charles}$ is $-\mathbb{I}\oplus-\mathbb{I}$ which means that $W=\Omega=-\mathbb{I}$ (here $\mathbb{I}$ reduces to the natural number ``1''). Under the transformation of this group element, there is an exchange symmetry between a particle with one unit of gauge charge and a particle with two units of gauge charge since the latter is trivially equivalent to a particle with gauge charge $-1$. There is also an exchange symmetry between a loop with magnetic flux $2\pi/3$ and a loop with magnetic flux $4\pi/3$ ($=-\frac{2\pi}{3}+2\pi$). These two exchange processes must occur  \emph{simultaneously}.

 \begin{table}
\caption{Examples of $\mathsf{Charles}$ (Sec.~\ref{sec:general_charlos}) when the matrix $K$ reduces to an integer.  A generic group element is denoted by $\mathscr{G}=W\oplus \Omega$. $\mathscr{G}_{\mathbb{I}}$ denotes the identity element: $\mathscr{G}_{\mathbb{I}}=\mathbb{I}\oplus \mathbb{I}$.  }\label{table:charlos}

 \begin{tabular}[t]{cc}
\hline

\hline
\hline
 \begin{minipage}[t]{1.1in}$K$\end{minipage} &\begin{minipage}[t]{2.1in} $\mathsf{Charles}$\end{minipage}     \\\hline
   \begin{minipage}[t]{0.64in} $K=\pm 1$ \end{minipage}&\begin{minipage}[t]{2in} $\{\mathscr{G}_{\mathbb{I}}\}$  \end{minipage}\\
\hline
   \begin{minipage}[t]{0.6in} $K=\pm 2$ \end{minipage}&\begin{minipage}[t]{2in} $\{\mathscr{G}_{\mathbb{I}}\}$ \end{minipage}\\
\hline
   \begin{minipage}[t]{0.6in} $|K|\geq 3$ \end{minipage}&\begin{minipage}[t]{2in}$\{\mathscr{G}_{\mathbb{I}},-\mathbb{I}\oplus-\mathbb{I}\}$ \end{minipage}\\
          \hline\hline
          
          \hline
  \end{tabular}
  \end{table}

  A simple example of a rank-2 $K$ matrix is $K=2\sigma_x$. If we do not worry about $\mathsf{Charles}$ for a moment, a diagonalization can be achieved by using $W=\sigma_x, \Omega=\mathbb{I}_{2\times 2}$. In the new basis, we end up with two copies of the level-2 topological $BF$ field theory,  thereby obtaining a $\mathbb{Z}_2\times\mathbb{Z}_2$ discrete gauge theory (i.e., $\mathbb{Z}_2\times\mathbb{Z}_2$  topological order). Due to Definition~\ref{dfn_charlos}, such a basis change is clearly not a group element of $\mathsf{Charles}$, but it reveals that the gauge structure is  $\mathbb{Z}_2\times\mathbb{Z}_2$ rather than $\mathbb{Z}_4$. It is important to distinguish these possibilities since those two gauge structures produce  the same ground state degeneracy ($\mathsf{GSD}$) on  a 3-torus \cite{bf1,bf2,bf3,horowitz89}. For this example, a typical group element of $\mathsf{Charles}$ is: $\mathscr{G}= \sigma_x\oplus \sigma_x$, which satisfies condition (i) in Eq.~(\ref{eqn:auto}).  
 Physically, $\Omega$ exchanges a particle labeled by $\mathbf{t}=(0,1)^T$ and a particle labeled by $\mathbf{t}=(1,0)^T$. At the same time, $\Omega$ exchanges  a loop labeled by $\mathbf{L}=(0,1)^T$ and a loop labeled by $\mathbf{L}=(1,0)^T$.

  For  convenience, condition (i) in Eq.~(\ref{eqn:auto}) can be visually represented by a tensor network-type graph  as shown in Fig.~\ref{figure_tensor}(a) of Appendix~\ref{appendix_proof_group}. It indicates that $K$ is a fixed point tensor (here, a matrix) that is invariant under $\mathsf{Charles}$   renormalization-group-like  transformations. The bond dimension is given by the rank of $K$. This graphical representation  allows us to straightforwardly generalize the notion of $\mathsf{Charles}$ to  more general Abelian topological quantum field theories (TQFTs) in (3+1)D that include more exotic topological terms. For instance, let us consider a TQFT with the  action:
  \begin{align} 
S=\frac{i}{2\pi}\int b^T\wedge K da+ i\int \Lambda^{IJK}a^I \wedge a^J  \wedge da^K\,,
\end{align} 
 where $\Lambda^{IJK}$ is a real tensor with three legs as shown in Fig.~\ref{figure_tensor}(b)  of Appendix~\ref{appendix_proof_group}.    By itself, and at a classical level,  the second term in this action corresponds to a topological invariant for the  mutual linkage of three electromagnetic flux loops \cite{bod}.  At  a quantum level, the action $S$ was also proposed as a continuum field theory description of Dijkgraaf-Witten lattice gauge theory \cite{dwitten,corbodism3}.  It can also be derived by gauging the global on-site symmetry group $G=\mathbb{Z}_{N_1}\times \mathbb{Z}_{N_2}\times \cdots$ of the TQFT action of a 3D SPT phase  with $N_1\times N_2\times \cdots=|\text{det}K|$, where the quantization of $\Lambda^{IJK}$ is determined by  the number of  topologically distinct ways  to impose $G$ in SPT phases \cite{YeGu2015}.
  The relation to  3-loop statistics \cite{wang_levin1,levin_talk,wang_levin2,lin_levin} is being investigated  \cite{YeGu2015,wang_levin1,levin_talk,3loop_ryu}. It is believed that the coefficient $\Lambda^{IJK}$ encodes the information of $3$-loop statistics that   classifies topologically distinct twisted discrete Abelian gauge field theories in (3+1)D. In analogy to the topological $BF$ field theory,  the   tensor $\Lambda^{IJK}$ must also be transformed accordingly under the charge-loop-lattice point group transformations. In order to keep the important 3-loop statistics data \cite{wang_levin1,levin_talk} invariant,  the generalized $\mathsf{Charles}$ should incorporate the following new condition:
\begin{align}
\sum_{I'J'K'}W^{I'I}W^{J'J}W^{K'K}\Lambda^{I'J'K'}=\Lambda^{IJK}\label{eqn:auto33}
\end{align}
 in addition to those conditions in Definition~\ref{dfn_charlos}. Likewise, we can also use a tensor-network-type graph [Fig.~\ref{figure_tensor}(b)  of Appendix~\ref{appendix_proof_group}], to graphically represent Eq.~(\ref{eqn:auto33}), where the bond dimension is no less than two.  
 
  Finally, we can also consider a TQFT with the   action: 
\!\!\! \begin{align}S=\frac{i}{2\pi}\!\int \!b^T\!\wedge K da+ i\!\!\int \!\Xi^{IJKL}a^I \wedge a^J  \wedge a^K\wedge a^L,
 \end{align}
where the coefficient $\Xi^{IJKL}$ is a tensor with four legs. Likewise, the quantized values of $\Xi$ encode the information of the four-loop braiding process \cite{wang_levin2}   and can provide topological invariants for classifying   twisted discrete gauge field theories in (3+1)D. In order to keep $\Xi$ invariant under point-group transformations of the charge-loop-lattice, the following relation should be obeyed:
 \begin{align}
\sum_{I'J'K'L'}\!W^{I'I}W^{J'J}W^{K'K}W^{L'L}\Xi^{I'J'K'L'}=\Xi^{IJKL}.\label{eqn:auto44}
\end{align}
A tensor-network-type representation is shown in Fig.~\ref{figure_tensor}(c)  of Appendix~\ref{appendix_proof_group}, where the bond dimension is no less than four.  

We may also consider the scenario that  $\mathsf{Charles}$ transformations can be performed \emph{locally} so that $\mathsf{Charles}$ becomes dynamically \emph{gauged}. In this case, the extrinsic twist defects become well-defined, deconfined excitations of a new topological phase.  
The resulting phases have been thoroughly studied in 2D and are non-Abelian topological phases called twist liquids \cite{Teo2015,Teo2014,ran,barkeshli_wen,Teo2013,Barkeshli2014,Bombin2010,genon_1,genon_2,genon_3,genon_4,You2012}. As a result, $W$ and $\Omega$ become space-time dependent. The difference between the next-nearest lattice sites is compensated by locally twisting matter fields. The tensor-network graph representations of the various symmetry transformations in Fig.~\ref{figure_tensor}  of Appendix~\ref{appendix_proof_group} are suggestive that such a tensor-network analysis may be a useful tool for future studies of 3D twist liquids.        
 
     \begin{figure}[t]
\centering
\includegraphics[width=8.5cm]{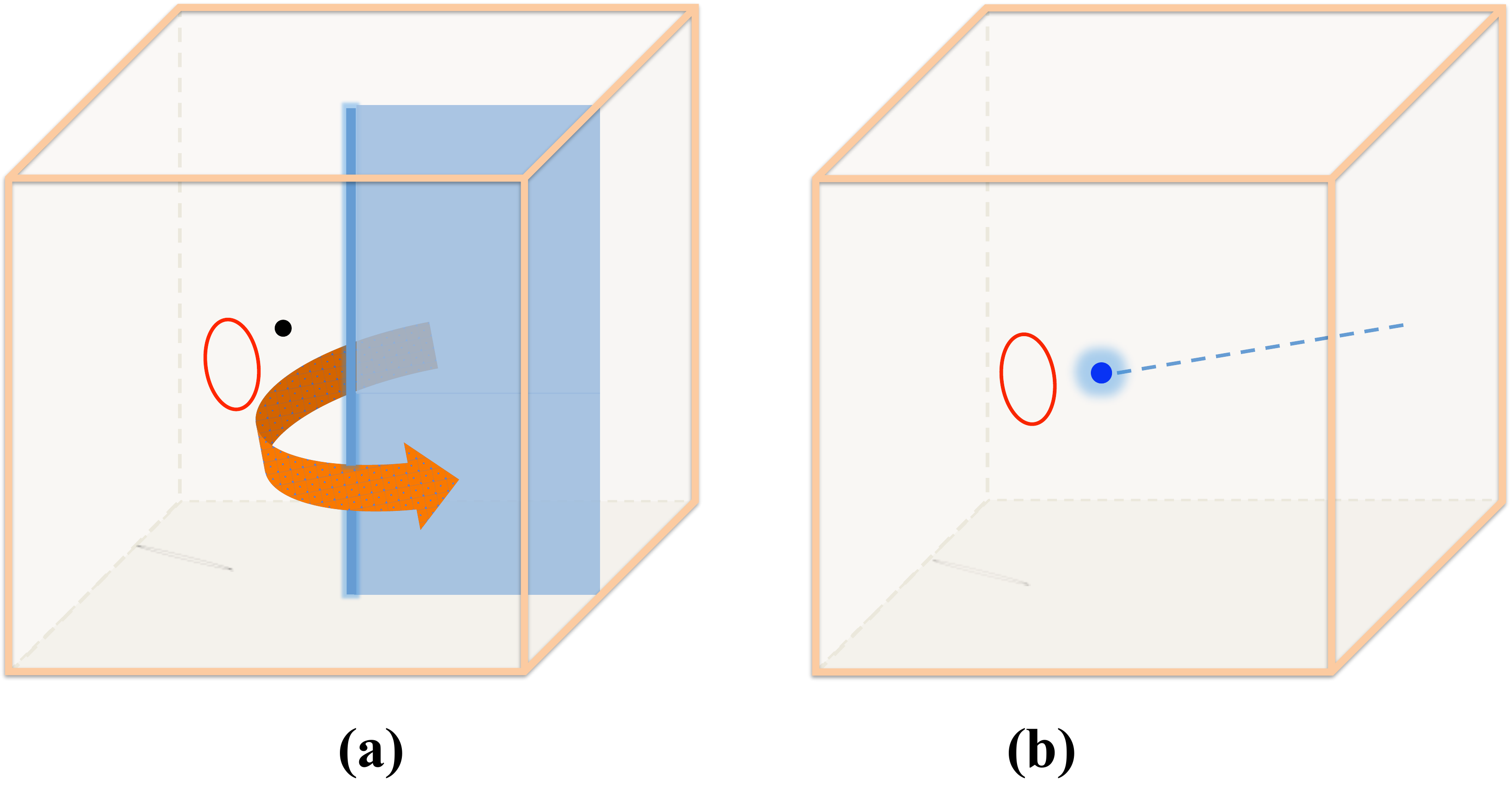}
\caption{(Color online) Extrinsic twist defects in 3D. (a) line defect; (b) point defect. The two cubic boxes denote the 3D bulk of an underlying quantum many-body system. The shaded plane in (a) denotes  a 2D branch cut/plane ending at the line defect, while the dashed line in (b) denotes a 1D branch cut ending at the point defect.      A line defect can act on both composite particles denoted by a black dot,  and loops denoted by a red circle.  Once   a point-like excitation and a loop excitation move around  a line defect,  the defect performs  the $\mathsf{Charles}$ symmetry transformation $\Omega$ and $W$ on the point-like excitation and the loop excitation, respectively.  In (b), the loop moves around the point defect such that the branch line intersects at the loop's spatial trajectory (a torus) only once. A $\mathsf{Charles}$ transformation induced by a point defect can only be $\mathscr{G}=W\oplus\Omega=\mathbb{I}\oplus\Omega$, which acts only on the loop excitations. However, due to Eq.~(\ref{eqn:auto}), the only candidate for $\Omega$ is $\mathbb{I}$. This means that  point defects  can only behave like the identity element $\mathscr{G}_{\mathbb{I}}$ of $\mathsf{Charles}$. Therefore, in 3D, we only consider line defects.}
\label{figure_defect}
\end{figure}

  \subsection{Theory of $\mathsf{Charles}$-defects: twist defect species and fusion}\label{sec:defect_fusion}

   \begin{figure*}[t]
\centering
\includegraphics[width=17.5cm]{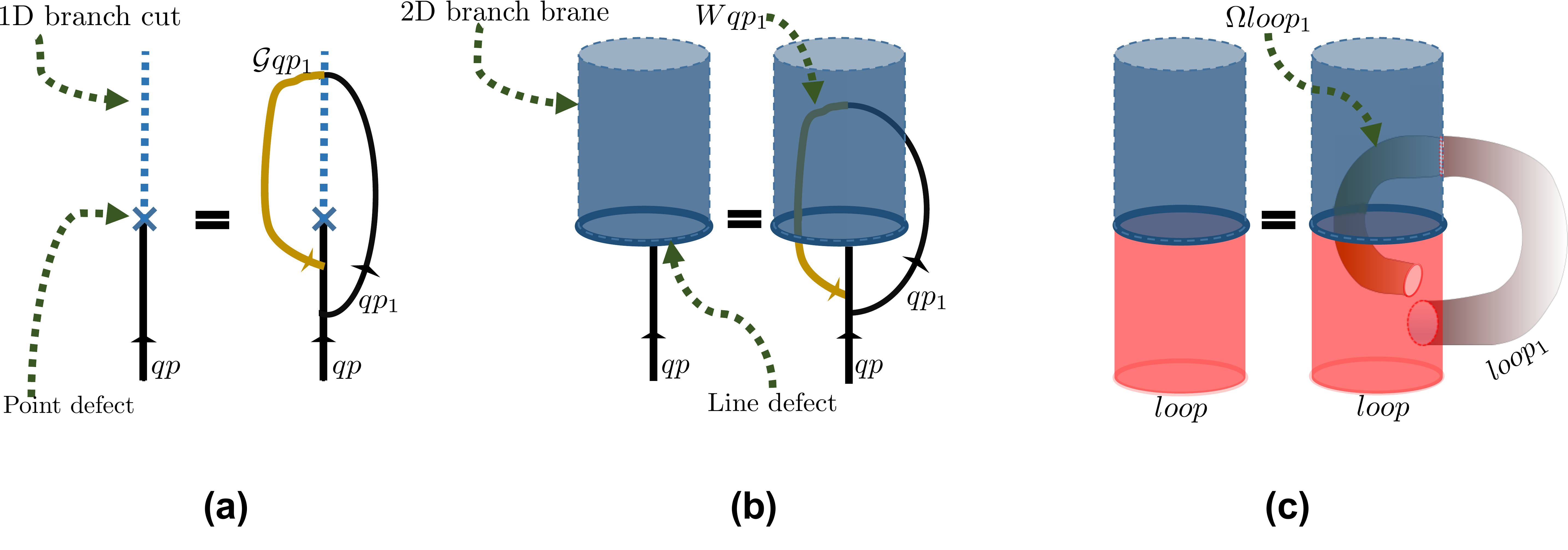}
\caption{(Color online) Diagrammatic description of equivalence classes of fusion rules. (a) shows the equivalent defect-anyon composites in 2D Abelian topological phases \cite{Teo2014}.  The bare point defect is labeled by a group element $\mathscr{G}$ of  the anyonic symmetry group. The dashed lines are branch cuts that end at the defect.  The vertical solid lines are quasiparticle (i.e., anyon, denoted by $qp$) strings (e.g., a string operator in the Wen-plaquette model), ending at the point defect. An anyon is transformed to another anyon when passing through the branch cut ($qp_1\rightarrow \mathscr{G}qp_1$). The fusion of a bare defect  denoted as $\mathcal{D}^0_\mathscr{G}$ and anyon ($qp$) forms a defect-anyon composite denoted as ``$\mathcal{D}^0_\mathscr{G} \times qp $''. It is topologically equivalent to the defect-anyon composite denoted as  $\mathcal{D}^0_\mathscr{G}\times 
(qp+(\mathbb{I}-\mathscr{G})qp_1)$ where $ qp_1$ denotes all   anyons of the 2D Abelian topological phase.
   In 3D systems where $\mathsf{Charles}$ replaces the anyonic symmetry of 2D systems, (b) and (c) show line defects (denoted as solid blue circles) on which the 2D branch branes (denoted as the surface of a cylinder) terminate. By passing through the 2D branch branes, a particle and a loop are transformed to another particle and loop, respectively ($qp_1\rightarrow W qp_1$, $loop_1\rightarrow \Omega loop_1$).
    In (b), the defect-charge (i.e., $qp$) composite denoted  $\mathcal{D}^0_{\mathscr{G}}\times qp$ 
is topologically equivalent to $\mathcal{D}^0_{\mathscr{G}}\times (qp+(\mathbb{I}-W)qp_1)$ where $ qp_1$ denotes all topologically distinct charge excitations.
 In (c), the red cylinder denotes the membrane operator that creates loop excitations and end on the line defect (the solid blue circles).  The defect-loop composite denoted  $\mathcal{D}^0_{\mathscr{G}}\times loop$
is topologically equivalent to $\mathcal{D}^0_{\mathscr{G}}\times (loop+(\mathbb{I}-\Omega)loop_1) $ where $\forall loop_1$ denotes  all topologically distinct loop excitations.     
} 
\label{figure_defect_composite}
\end{figure*}

In the following, we study  extrinsic twist defects associated with $\mathsf{Charles}$ group elements in analogy to  extrinsic twist defects in 2D Abelian topological phases with anyonic symmetry \cite{Teo2015,Teo2014,Teo2013,Barkeshli2014,genon_1,genon_2,Bombin2010,You2012}. More specifically, we  explore  two issues: (i)  the universal labeling of a defect in 3D, and (ii)   the fusion properties of defect-charge-loop composites. Recently  remarkable progress in the study of various aspects of string/loop excitations in (3+1)D topological phases of matter, such as their description using lattice models and field theories, their associated ground-state degeneracy ($\mathsf{GSD}$), and their braiding and fusion properties has been made\cite{string1,string1.5,wang_levin1,string2,string3,string4,ran,string5,string6,string7,string8,string9,string10}.  As will be seen below, the consideration of extrinsic defects within the framework of charge-loop excitation symmetry introduces a new aspect to the physics of loop excitations in (3+1)D topological phases.

We begin by reviewing the physics of twist defects and defect-anyon composites in 2D Abelian topological phases. From a topological point of view, externally imposed defects in such phases are  a set of special, potentially non-Abelian  objects. For Abelian groups, each group element of the anyonic symmetry group $G$ corresponds to a bare defect and there are $\mathsf{ord}(G)$ distinct bare defects, where   $\mathsf{ord}(G)$ is the order of $G$.  Generically, the defects are labeled by the conjugacy classes of $G,$ but since we only deal with Abelian groups here we will not often make this distinction.   By a ``bare'' defect, we really mean that the defect is externally imposed alone in the bulk.  
In general, defects can be bound to anyons of the parent Abelian topological phase,   thereby forming a \emph{defect-anyon composite} which is, by definition, not bare. When given a group element of $G$, the total number  of topologically distinct defects  (also known as defect species) includes bare ones and composite ones, and is not always the same as the number of topologically distinct anyons. In other words, two defect-anyon composites might be topologically equivalent to each other if  there do not exist gauge-invariant Wilson measurements that can distinguish them. Indeed, there is a consistency equation for determining the equivalence classes of defect types \cite{Teo2014,Teo2015},  
\begin{align}
\mathcal{D}^0_\mathscr{G} \times qp =\mathcal{D}^0_\mathscr{G}\times 
(qp+(\mathbb{I}-\mathscr{G})qp_1) \,,\label{equation_fusion_2d}\end{align}
which is diagrammatically shown in Fig.~\ref{figure_defect_composite}(a). Here, $qp$ denotes a quasiparticle (i.e., an anyon) that is provided by the parent 2D Abelian topological phase. $\mathcal{D}^0_\mathscr{G}$ denotes a bare defect labeled by a  group element $\mathscr{G}$ of the anyonic symmetry group $G$.   The particle $qp_1$ is any anyon provided by  the parent 2D Abelian topological phase.
The composite object $ \mathcal{D}^0_{\mathscr{G}}\times qp$ denotes the fusion between $\mathcal{D}^0_{\mathscr{G}}$ and $qp$ that forms a defect-anyon composite.  Specifically, Eq.~(\ref{equation_fusion_2d}) determines when this defect-anyon composite is topologically identical to a defect-anyon composite that is formed by the fusion between the same bare defect and a different anyon given by $qp+(\mathbb{I}-\mathscr{G})qp_1$.  The symbol ``$+$'' should be regarded as the  addition of quasiparticle vectors in the $K$-matrix Chern-Simons theory. The physical reason of this equivalence is really due to the nontrivial internal structure of   a defect-anyon composite. More specifically, the anyon $qp$ that is trapped at the defect can emit anyon $qp_1$ which moves around the defect once. As a result, anyon $qp_1$ is changed to anyon $\mathscr{G}qp_1$ that is finally absorbed by the defect. Such a process occurs inside the defect-anyon composite and cannot change the defect species \cite{Teo2014}.  Therefore, the process provides an equivalence between two defect-anyon composites.

  A typical example in (2+1)D is $K=2\sigma_x$ Chern-Simons theory that describes $\Z_2$ topological order. Its anyonic symmetry group is given by $\{\mathbb{I},\sigma_x\}$. The nontrivial group element $\sigma_x$ interchanges  the anyon $e$,  labeled by the quasiparticle vector $(1 \text{ mod }2,0\text{ mod }2)^T$, and the anyon $m$, labeled by the quasiparticle vector $(0\text{ mod }2,1\text{ mod }2)^T$. A defect labeled by this group element can in principle be realized by externally imposing a dislocation in the Wen-plaquette model as mentioned previously.  For   convenience, the identity quasiparticle (vacuum) $vac$ is labeled by $(0\text{ mod }2,0\text{ mod }2)^T,$ and the fermion quasiparticle $\psi$ is labeled by   $(1\text{ mod }2,1\text{ mod }2)^T$. Thus, the only non-trivial, bare defect is given by $\mathcal{D}^0_{\sigma_x}$. Next we need to deduce equivalence classes of defect-anyon composites. Taking  into account Eq.~(\ref{equation_fusion_2d}) and  $\mathbb{I}-\sigma_x=\left(\begin{smallmatrix}1&-1\\-1&1\end{smallmatrix}\right)$,  we have: 
  \begin{align}
  &\mathcal{D}^0_{\sigma_x}\times qp=\mathcal{D}^0_{\sigma_x}\times (qp+\epsilon)\,,  
  \forall qp\in\{vac,e,m,\epsilon\}.\nonumber
  \end{align}  Here, the symbol ``$+$'' denotes the usual addition of quasiparticle vectors of $qp$ and $\epsilon$. Therefore, there are two equivalence classes:   $\mathcal{D}^0_{\sigma_x}\times e=\mathcal{D}^0_{\sigma_x}\times m$ and $\mathcal{D}^0_{\sigma_x}\times \psi=\mathcal{D}^0_{\sigma_x}$. In other words, there are two topologically distinct defects: one is bare, given by a bare defect $\mathcal{D}^0_{\sigma_x}$; the other one is a defect-anyon composite, denoted by $\mathcal{D}^1_{\sigma_x}=\mathcal{D}^0_{\sigma_x}\times e$.  The fusion rules of these two defects are given by: 
  \begin{align}
  \mathcal{D}^0_{\sigma_x}\times\mathcal{D}^0_{\sigma_x}=\mathcal{D}^1_{\sigma_x}\times\mathcal{D}^1_{\sigma_x}=vac+\psi, \mathcal{D}^0_{\sigma_x}\times\mathcal{D}^1_{\sigma_x}=e+m\,,\nonumber\end{align} where ``$+$'' here denotes the collection of different fusion channels into quasiparticles of simple-type.

Now that we have reviewed the lower dimensional case, let us move on to 3D.  Simply from a dimensionality point of view, there are two types of extrinsic defects in 3D: line defects and point defects. The latter also appear in 2D and serve as  end points on which 1D branch cuts (i.e., the dashed line in Fig.~\ref{figure_defect}(b)) terminate. The former are really loop-like. In Fig.~\ref{figure_defect}(a), the line defect is drawn as a finite line that ends at the top and bottom boundaries where a periodic boundary condition is implicitly imposed. A 2D branch ``brane'' (i.e., the shaded plane in Fig.~\ref{figure_defect}(a)) is attached to each line defect.

From Fig.~\ref{figure_defect}, we see that line defects can perform generic $\mathsf{Charles}$ operations where both point particles and loops are transformed. In contrast, point defects can only perform $\mathsf{Charles}$ operations on loops,  meaning that $\mathscr{G}=W\oplus \Omega=\mathbb{I}\oplus \Omega$ for point defects.   However, due to Eq.~(\ref{eqn:auto}), the only candidate for $\Omega$ is $\mathbb{I}$. This means that  point defects can only behave like the identity element $\mathscr{G}_{\mathbb{I}}=\mathbb{I}\oplus\mathbb{I}$ of $\mathsf{Charles}$. Therefore, in 3D, we only consider line defects since point defects \emph{cannot} perform nontrivial $\mathsf{Charles}$ operations.

In a   manner similar to 2D, a \emph{defect-charge-loop composite} is allowed, where the term ``defect'' corresponds to  a line defect,  ``charge''  corresponds to  a point-like excitation,  and ``loop''  corresponds to a loop excitation. Since loops are always transformed to loops by $\Omega,$ and particles are always transformed to particles by $W$,  we may study defect-charge-composites and defect-loop-composites separately. In order to determine  defect species for a given $\mathsf{Charles}$ group element $\mathscr{G}$, we need to study the equivalence classes of the above two kinds of defect composites.  For defect-charge-composites, the following equation determines the equivalence classes:
\begin{align}
\mathcal{D}^0_{\mathscr{G}}\times qp=\mathcal{D}^0_{\mathscr{G}}\times (qp+(\mathbb{I}-W)qp_1)\,,\label{equation_fusion_3d_charge}
\end{align}
which is diagrammatically shown in Fig.~\ref{figure_defect_composite}(b).   $qp$ denotes point-like particle excitations.  $\mathcal{D}^0_{\mathscr{G}}$ denotes the bare line defect that is  labeled by a $\mathsf{Charles}$ group element (or conjugacy class for a non-Abelian group) $\mathscr{G}=W\oplus \Omega$.     $  qp_1$  is any particle excitation provided by the parent 3D Abelian topological phase. Eq.~(\ref{equation_fusion_3d_charge}) means that the defect-charge composite  $\mathcal{D}^0_{\mathscr{G}}\times qp$ is topologically equivalent to the defect-charge composite that is formed by the fusion between the same bare line defect and a different particle excitation given by $(qp+(\mathbb{I}-W)qp_1)$.  
Likewise, we have a similar equation for defect-loop composites:
\begin{align}
\mathcal{D}^0_{\mathscr{G}}\times loop=\mathcal{D}^0_{\mathscr{G}}\times (loop+(\mathbb{I}-\Omega)loop_1)\,,\label{equation_fusion_3d_loop}
\end{align}
which is diagrammatically shown in Fig.~\ref{figure_defect_composite}(c).  One can also unify Eqs.~(\ref{equation_fusion_3d_charge},\ref{equation_fusion_3d_loop}) by considering charge-loop composites. We will show this in the following example.

Let us take $K=3$ in Table~\ref{table:charlos} as an example. There is only one nontrivial group element given by $\mathscr{G}=-\mathbb{I}\oplus-\mathbb{I}$. For   convenience, we label the three topologically distinct particle excitations as $t_0,t_1,t_2$ and the three distinct loop excitations as $l_0,l_1,l_2$. Using numerical labels, we have: $t_0=0\text{ mod }3, t_1=1\text{ mod }3, t_2=2\text{ mod }3$, and $l_0=0\text{ mod }3, l_1=1\text{ mod }3, l_2=2\text{ mod }3$. We can consider the set of 2D vectors $\mathbf{V}_{ij}=(t_i,l_j)^T$ where $i,j=0,1,2$ and hence,  there are $3^2=9$  vectors that label the 9 topologically distinct charge-loop composites:
\begin{align}
\!\!\!\!\{\mathbf{V}_{ij}\}\!=\!\!\bpm0 \\ 0\epm\!\!, \bpm1 \\ 0\epm\!\!, \bpm2 \\ 0\epm\!\!, \bpm0 \\1\epm\!\!, \bpm1\\1 \epm\!\!, \bpm2 \\ 1\epm\!\!, \bpm0 \\ 2\epm\!\!,\bpm1 \\ 2\epm\!\!, \bpm2\\ 2\epm.\nonumber
\end{align}  As a result,   Eqs.~(\ref{equation_fusion_3d_charge},\ref{equation_fusion_3d_loop}) can be unified as:
\begin{align}
\mathcal{D}^0_{\mathscr{G}}\times \mathbf{V}=\mathcal{D}^0_{\mathscr{G}}\times (\mathbf{V}+(\mathscr{G}_{\mathbb{I}}-\mathscr{G})\mathbf{V}')\,,\label{equation_fusion_3d_unified}
\end{align}
where $\mathbf{V},\mathbf{V}'\in\{\mathbf{V}_{ij}\}$.  By noting that 
$\mathscr{G}_{\mathbb{I}}-\mathscr{G}=\left(\begin{smallmatrix}1&0\\ 0&1\end{smallmatrix}\right)-\left(\begin{smallmatrix}-1&0\\ 0&-1\end{smallmatrix}\right)=\left(\begin{smallmatrix}2&0\\ 0&2\end{smallmatrix}\right)$,   the above relation reduces to: 
\begin{align}
\mathcal{D}^0_{\mathscr{G}}\times \mathbf{V}=\mathcal{D}^0_{\mathscr{G}}\times (\mathbf{V}+2\mathbf{V}').
\end{align} 
As a result, all defect-charge-loop composites are topologically equivalent to the bare defect:
$  \mathcal{D}^0_{\mathscr{G}}=\mathcal{D}^0_{\mathscr{G}}\times \mathbf{V}_{ij}\,,$ 
where $\{\mathbf{V}_{ij}\}$ denotes the 9 vectors ($i,j=0,1,2$).  
 The resulting fusion rules are given by:
 \begin{align}
 \mathbf{V}_{ij}\times  \mathbf{V}_{i'j'}&=\mathbf{V}_{(i+i')\text{mod}3,(j+j')\text{mod}3}\,,\\
   \mathcal{D}^0_{\mathscr{G}}\times   \mathcal{D}^0_{\mathscr{G}}&=\sum_{ij} \mathbf{V}_{ij}\,,
 \end{align}
 from which we see that there are multiple  fusion channels when two defects are fused together. It indicates that the externally imposed line defect $\mathcal{D}^0_{\mathscr{G}}$ is of non-Abelian nature.


 \section{Conclusions}\label{sec:conclusion_direction}

     In this work, a composite particle theory for 3D fermionic gapped phases was formulated based on a specific parton construction of electrons.  Composite particles are  bound states of partons and magnetic monopoles for a set of internal gauge fields and the external electromagnetic field $A_\mu$. The resulting fully-gapped phases were constructed by condensing two composite particles. All excitations including point-like and string-like excitations as a whole form a charge-loop-lattice. Each site of the charge-loop-lattice corresponds to a deconfined excitation of the condensed phase. A general mechanism for charge fractionalization in 3D was studied in detail. Based on the general framework of   composite particle theory, we further explored two important properties of 3D Abelian topological phases.  First, we studied phases with non-vanishing axion $\Theta$ angle which is characteristic of the tilted charge lattice. It was found  that time-reversal invariant fractional topological insulators with $\Theta\neq \pi$ can be constructed from   composite particle theory. 
     Second, we generalized the notion of anyonic symmetry of 2D Abelian topological phases to a charge-loop excitation permutation symmetry ($\mathsf{Charles}$) group in 3D Abelian topological phases. We also investigated the relation between $\mathsf{Charles}$ group elements and line twist defects in (3+1)D Abelian topological phases.
 
     There are several interesting directions for future studies. 
          \emph{First}, it is interesting to propose a systematic theory of the symmetric surface states of fractional topological insulators based on the composite particle theory. The 2D surface may exhibit quantum phenomena that are even more exotic than the surface topological order recently found on the surface of interacting topological insulators and interacting bosonic topological insulators  \cite{bti1,bti2,bti3,bti4,bti5,bti7,bti8,sto1,sto2,sto3,sto4,sto5}.  For the {FTI}  bulk lattice model construction and the phase diagram of confinement-deconfinement,  the idea in Ref.~\onlinecite{kb2015} may be helpful.  
              \emph{Second}, one may consider the composite particle theory by assuming that partons form topological superconductor ans\"atze,  which may lead to interacting topological superconductors with fractional gravito-electromagnetism and a fractional version of the gravitational Witten effect \cite{grav1,grav2}. 
           \emph{Third}, as discussed in Sec.~\ref{sec:general_charlos}, the tensor-network-type graphs may be helpful for understanding 3D analogs of the  twist liquid, i.e., the topological phases obtained by gauging $\mathsf{Charles}$.           
               \emph{Fourth}, it is interesting to think if there are simple 3D lattice models that can demonstrate the physics of extrinsic defects and $\mathsf{Charles}$, in analogy to the 2D case where there are lattice models like the Wen-plaquette model.  In addition, some group elements of $\mathsf{Charles}$ may break $U(1)$ charge symmetry. A line defect associated with such a group element might be realized in a $U(1)$-symmetric 3D lattice model as an extrinsic defect coated with a superconducting region. 
          \emph{Fifth},  in analogy to 2D anyonic symmetry where $G$-crossed tensor category theory \cite{Teo2015,Barkeshli2014} was proposed, a generic mathematical framework is also needed for 3D extrinsic defects. 
            \emph{Sixth},  it would be useful to have a microscopic theory of 3D line twist defects in terms of a cutting and gluing procedure where the twist defects are formed by tuning/twisting allowed tunneling terms between the two sides of a gapless cut \cite{genon_1,genon_2,genon_3,dhl_cut_glue}.

 \section*{Acknowledgement}
We would like to thank K.~Shiozaki, Y.~Qi, and S.~Ryu for helpful discussions. P.Y. would like to thank  Z.-Y.~Weng and X.-G.~Wen for beneficial collaborations and insightful discussions on parton constructions, and also acknowledges S.-T.~Yau's  hospitality  at the Center of Mathematical Sciences and Applications at Harvard University where the  work was done in part.   This work was  supported in part by the NSF through grant DMR 1408713 at the University of Illinois.(P.Y. \& E.F.)  T.L.H. is supported by the US National
Science Foundation under grant DMR 1351895-CAR. J.M. was supported by NSERC grant \#RGPIN-2014-4608, the CRC Program, CIFAR, and the University of Alberta.

\appendix
\section{Summary of notations, abbreviations, and definitions}\label{appendix:notation}
In this Appendix, several notations, abbreviations,   definitions, and criteria are collected for the reader's convenience.  
 
 \bigskip

 \emph{1. Mathematical notations:}

\bigskip

 $u,v,u',v',q,q'$: a set of parameters that label the two condensed composites as shown in Table. \ref{table:em}.

 $Q$: the net EM electric charge carried by a composite.

 $Q_{\rm Debye}$: the screening charge cloud around a composite. It is induced by the two composite condensates $\varphi_1$ and $\varphi_2$.

 $N_A$: the bare EM electric charge carried by a composite. It is related to $Q$ via Eqs.~(\ref{NE1},\ref{NE_screen}).

 $M$: the EM magnetic charge carried by a composite.

 $M_\mu$: the 4-current of EM magnetic monopoles, introduced in Sec.~\ref{sec_FTI_Coulomb}.

 $N_{a,b}$: gauge charges in $U(1)_a$ and $U(1)_b$ gauge groups. An integer vector $\mathbf{N}_e$ is formed via Eq.~(\ref{define_KNq}).

 $N_m^{a,b}$: magnetic charges in $U(1)_a$ and $U(1)_b$ gauge groups.  An integer vector $\mathbf{N}_m$ is formed via Eq.~(\ref{define_KNq1}).

$\Gamma$: self-statistics of a composite. $\Gamma$ is even (odd) if the composite is bosonic (fermionic), see Eq.~(\ref{eq:stat123}).

 $\theta$: $\theta=0$ if all partons ($f^1,f^2,f^3$) form trivial band insulators. $\theta=\pi$ if all partons form topological insulators.
 
 $\vartheta^{cl}$: the mutual statistics between a point-like particle excitation and a loop excitation, see Eq.~(\ref{eq:cltheta}).
 
  $\Theta$: the axion angle of the electron states (i.e., the resulting fermionic gapped phase constructed via the composite particle theory).
 
  $g_{a,b}$ dimensionless gauge coupling constants of $U(1)_{a,b}$ gauge groups.

 $\mathcal{D}^0_\mathscr{G}$: a bare line defect associated with $\mathsf{Charles}$ group element $\mathscr{G}$.
 
 \bigskip

\emph{2. Abbreviation:}

\bigskip

 $\mathsf{Charles}$: \textbf{char}ge-\textbf{l}oop \textbf{e}xcitation \textbf{s}ymmetry.
 
 EM: electromagnetic (specific to the usual background electromagnetic  field $A_\mu$).

  FQH: fractional quantum Hall effect.

 {FTI}: fractional topological insulator.

 $\mathsf{GCD}$: greatest common divisor.

  GSD: ground state degeneracy.

 IQH: integer quantum Hall effect.

 SET: symmetry-enriched topological phase.

 SPT: symmetry-protected topological phase.

 TI: free-fermion topological insulator.

 TO: topological order

 TQFT: topological quantum field theory.
 
 \bigskip

\emph{3. Definitions:}

\bigskip

 Loop-lattice:  Definition~\ref{dfn_loops} on Page~\pageref{dfn_loops}.

 Excitation and charge lattice: Definition~\ref{dfn_excitation} on Page~\pageref{dfn_excitation}.

Intrinsic excitation and intrinsic charge lattice: Definition~\ref{dfn_intrinsic_exc} on Page~\pageref{dfn_intrinsic_exc}. 

 Charge-loop-lattice: Definition~\ref{dfn_cloop} on Page~\pageref{dfn_cloop}.

 Charge-loop excitation symmetry: Definition~\ref{dfn_charlos} on Page~\pageref{dfn_charlos}.

 \bigskip

\emph{4. Others}

\bigskip

Criterion~\ref{crt_loop_exc} for loop excitations on Page \pageref{crt_loop_exc}

Criterion~\ref{crt_charge_1} for charge fractionalization on Page \pageref{crt_charge_1}

Criterion~\ref{crt_charge_2} for charge fractionalization on Page \pageref{crt_charge_2}

 \section{Technical details in Sec.~\ref{sec:gauge_strc}}
 \subsection{Details of Eq.~(\ref{constr_3})}\label{appendix_bosonic}
\begin{widetext}

By inserting the data of $\varphi_1$ and $\varphi_2$ in Table~\ref{table:em} into Eq.~(\ref{eq:quantum_stat}), one may obtain:
\begin{align}
 \Gamma(\varphi_1)
= 3q-\frac{\theta}{2\pi}[u(u+1)+v(v+1)+(u+v)(u+v-1)]\,,
\end{align}
 and
 \begin{align}
 \Gamma(\varphi_1)   
 =3q'-\frac{\theta}{2\pi}[u'(u'+1)+v'(v'+1)+(u'+v')(u'+v'-1)].
 \end{align}  
\end{widetext}
 

 \subsection{Proof of Theorem \ref{theorem_flux}}\label{appendix_proof_theorem_flux}
We present B\'{e}zout's lemma  as a preliminary:
\begin{description}
\item [B\'{e}zout's lemma \cite{bezout79}] Let $a$ and $b$ be nonzero integers and let $d$ be their greatest common divisor ($\mathsf{GCD}$). Then there exist integers $x$ and $y$ such that  $ax+by=d$. 
In addition,   $d$ is the smallest positive integer that can be written as $ax + by$; 
every integer of the form $ax + by$ is a multiple of  $d$. 
\end{description}

Let us now prove Theorem 1. 

\begin{proof}

 {Sufficiency:}   When $|uv'-u'v|=1$, according to Eq.~(\ref{eq:ineq}), we straightforwardly obtain  $|\mathsf{GCD}(u,u')|=1,|\mathsf{GCD}(v,v')|=1$. Then, the equalities in Eq.~(\ref{eq:ineq}) hold. Therefore,   $(\Phi^a_e)_{\rm min}=2\pi, (\Phi^b_e)_{\rm min}=2\pi$ in Eq.~(\ref{discreteflux1}).

 {Necessity:} We start with the equalities in Eq.~(\ref{eq:ineq}), i.e., $|uv'-u'v|=|\mathsf{GCD}(u,u')|=|\mathsf{GCD}(v,v')|$. If $|uv'-u'v|\neq1$, meaning that $|\mathsf{GCD}(u,u')|=|\mathsf{GCD}(v,v')|\neq 1$. Therefore, $u,u'$ are not co-prime; $v,v'$ are not co-prime. Then, we consider:
\begin{align}
1=\bigg|\frac{u}{\mathsf{GCD}(u,u')}v'-\frac{u'}{\mathsf{GCD}(u,u')}v\bigg|\,,\label{in1}
\end{align}
where $\frac{u}{\mathsf{GCD}(u,u')}, \frac{u'}{\mathsf{GCD}(u,u')}$ are co-prime by definition, i.e.,
\begin{align}
 \bigg|\mathsf{GCD}(\frac{u}{\mathsf{GCD}(u,u')}, \frac{u'}{\mathsf{GCD}(u,u')})\bigg|=1\,.\nonumber
 \end{align}
  Then, according to Eq.~(\ref{eq:ineq}), we can also have the following inequalities if we just replace $u$ and $u'$ in Eq.~(\ref{eq:ineq}) by $\frac{u}{\mathsf{GCD}(u,u')}$ and $\frac{u'}{\mathsf{GCD}(u,u')}$, respectively:
\begin{align}
\bigg|\frac{u}{\mathsf{GCD}(u,u')}v'-\frac{u'}{\mathsf{GCD}(u,u')}v\bigg|\geq |\mathsf{GCD}(v,v')|\,.\label{in2}
\end{align}
Due to Eqs.~(\ref{in1},\ref{in2}), we obtain $|\mathsf{GCD}(v,v')|=1$. This is contradictory to our starting point $|\mathsf{GCD}(v,v')|\neq 1$. 
Therefore, the only possibility is $|uv'-u'v|=|\mathsf{Det}K|=1$. 

\end{proof}


\subsection{Equivalence between Criterion~\ref{crt_charge_1} and Criterion~\ref{crt_charge_2}}\label{appendix_dirac}
Consider two excitations in a $U(1)_{\rm EM}$-symmetric system. Let one carry  zero EM magnetic charge $M=0$ and minimal non-vanishing EM electric charge $Q=\frac{1}{w}$ with $w\in\Z$.  Let the other excitation carry a minimal nonzero EM magnetic charge $w'$ and an EM electric charge, say, $y$. $y$ can be either integer or non-integer.  
Due to the Dirac-Zwanziger-Schwinger quantization condition \cite{ dirac1,dirac2,dirac2.5,dirac3,dirac3.5},   the magnetic and electric charges of the above two excitations satisfy:
\begin{align}
\left(\frac{1}{w}\times w'-0\times y\right)=0,\pm1,\pm2,\cdots\,.
\end{align}
Therefore, the minimal choice of $w'$ is $w'=w$, indicating that the change of   quantization of the EM magnetic charge $M$  is accompanied with a change of the charge quantization. Once $w>1$, $w'$ is also larger than one. In this sense, the two criteria are equivalent.

\subsection{Debye-H{\"u}ckel charge cloud $Q_{\rm Debye}$ is the unique source of   charge fractionalization}\label{appendix_na_integer}
 
In this Appendix, we prove that the Debye-H{\"u}ckel charge cloud $Q_{\rm Debye}$ defined in Eq.~(\ref{NE_screen}) is the \emph{unique} source of   charge fractionalization. In other words, $N_A$ is always integer-valued when $M=0$. By definition in Eqs.~(\ref{eqn:transition_matrix},\ref{eqn:transition_matrix1},\ref{Nnf1}), $N_A$ is given by:
\begin{align}
 N_A=&N^{f1}+N^{f2}-N^{f3}\nonumber\\
 =&(n^{f1}+n^{f2}-n^{f3})+\frac{3\theta }{2\pi}M-\frac{2\theta}{2\pi}N^b_m\,,
\end{align}
where $n^{fi}$ are integer-valued. $N^b_m$ is integer-valued, and $\theta=0,\pi$. Therefore, $-\frac{2\theta}{2\pi}N^b_m$ is always integer-valued. As a result, $N_A$ is integer-valued when $M=0$.
 

 \section{Technical details in Sec.~\ref{sec:theta}}

 \subsection{Partons occupying  trivial bands}\label{appendix_Gamma_trivial_1}
 We assume that all partons $f^{i}$ (pure gauge charge carriers) form three   trivial band insulators ($\theta=0$).   According to Eqs.~(\ref{constr_1plus},\ref{constr_1},\ref{constr_3}), we have:
\begin{align}
u,v,u',v'\in\Z\,;~~q,q'\in\Z_{\rm even} \,.\label{theta_0_q_even}
 \end{align}
 Since $\theta=0$, we have:
 \begin{align}
 N^{fi}=n^{fi}\in\Z
 \end{align}
 according to Eq.~(\ref{Nnf1}).    Due to the definitions in Sec.~\ref{sec:gauge_strc}, we end up with:
 \begin{align}N_a,N_b\in\Z\,.
 \end{align} 
  Thus, in the mean-field ans\"atze with $\theta=0$, all magnetic charges and electric charges of composites are integer-valued. However, $Q_{\rm Debye}$ and $Q$ may be fractional, depending on the condensate parameters.  
 
According to Definition~\ref{dfn_excitation}, excitations  are a subset of  generic composites and  satisfy the two equations in Eq.~(\ref{excitation_1}).  Thus, only a 4D sublattice embedded in the 6D lattice survives, i.e.,  the charge lattice in Definition~\ref{dfn_excitation}.  Since $N_a(=N^{f1}-N^{f2})$ and $N_b(=N^{f3}-N^{f2})$ are fully determined by $M$ via Eq.~(\ref{NabM}),   we may use the labels $(N^{f2},M,N^a_m,N_m^b).$ These four linearly independent integer numbers are ``4D coordinates'' of the 4D lattice that label excitations. Then, the bare EM electric charge $N_A$ is expressed as:
\begin{align}
N_A=N^{f1}+N^{f2}-N^{f3}=(r-s)M+N^{f2}\,.
\end{align}
The net EM electric charge $Q$ is given by:
\begin{align} 
Q=&N_A-Q_{\rm Debye}\nonumber\\
=&N^{f1}+N^{f2}-N^{f3}-Q_{\rm Debye}\nonumber\\
=&(r-s)M + N^{f2}-rN_m^a-sN_m^b\,,\label{q_trivial}
\end{align}
 where $N^{f1}-N^{f2}=rM$, $N^{f3}-N^{f2}=sM$ with $r$ and $s$:
 \begin{align}
 r=\frac{qv'-q'v}{\mathsf{Det}K}\,,~s=\frac{q'u-qu'}{\mathsf{Det}K}\,.
 \end{align}
 We note  that $r$ and $s$ can be either integer or non-integral rational numbers.  However, $rM$ and $sM$ must be integer-valued in order to ensure the $N^{fi}$ are integer-valued. Thus, the quantization of $M$ should be altered properly if $r$ and $s$ are non-integral rational numbers.   In summary, we can define the following domains:
 \begin{align}
N^{f2}\in\Z,N^a_m\in\Z,N^b_m\in\Z\,, \frac{M}{w}\in\Z\,,
\end{align}
where $w$ is a positive minimal integer such that both $rM\in\Z$ and $sM\in\Z$ are satisfied.  
In Eq.~(\ref{q_trivial}),  the $M$-dependent charge $(r-s)M$ is integer-valued:
 \begin{align}
 (r-s)M\in\Z\,.\label{theta_trivial_response}
 \end{align}
 Therefore, the minimal quantized value of $Q$ is sufficiently determined by $rN_m^a$ and $sN^b_m$ by noting that the latter two terms can be potentially fractionalized depending on $r$ and $s$.  In the language of the EM response theory,  $M$-dependent charge means that the EM magnetic current   minimally couples to the EM gauge field $A_\mu$. In other words, the bulk supports an EM response action with $\Theta$ term. If we define $\frac{\Theta}{2\pi}M=(r-s)M$, then $Q=\frac{\Theta}{2\pi}M+N^{f2}-rN_m^a-sN^b_m$ with $\Theta=2\pi(r-s)$. However, due to Eq.~(\ref{theta_trivial_response}), this nonzero $\Theta$ gives rise to an integer charge cloud surrounding EM magnetic monopoles. This additional charge cloud does not render a new quantization of $Q$  different from the quantization when $M=0$. In other words, the EM charge lattice ($M-Q$ plane) is just a square lattice that is not tilted. The allowed values of $Q$ when $M=0$ are completely the same as   when $M\neq 0$.   In this sense,  the resulting state with $\Theta=2\pi(r-s)$ is equivalent to a trivial state with $\Theta=0$. By comparison, a typical example with nontrivial $\Theta$ angle has a $Q$ quantization shown in Eqs.~(\ref{Q_fractional_M0},\ref{Q_fractional_M1}) of Sec.~\ref{sec_FTI} where the quantization of $Q$  manifestly depends on $M$.

   \subsection{Derivation of Eqs.~(\ref{Q_fractional_M0},\ref{Q_fractional_M1}), Eq.~(\ref{eq:stat4321_new}), and the site distribution in Fig.~\ref{figure_theta}(c)}\label{appendix_gamma_stat4321_new}
   
   Since $M$ is quantized in multiples of $3$ as indicated in Eq.~(\ref{double_m_charge}), one may introduce an integer $k$ such that $M=3k$. Meanwhile, Eq.~(\ref{Q_expression}) indicates that $Q$ is generically quantized in multiples of $1/6$. Thus, we can introduce an integer $k_0$ such that $Q=\frac{k_0}{6}$. Then,   Eq.~(\ref{Q_expression}) is formulated as:
\begin{align}
k_0+5k=2(7n^{f1}-n^{f2}-3n^{f3})\,,
\end{align}
where the r.h.s. is always even. $5k$ has the same even-odd property as $k$. As a result, $k_0$ and $k$ must be simultaneously either odd or even, which leads to Eqs.~(\ref{Q_fractional_M0},\ref{Q_fractional_M1}).

Then,  we start with $\Gamma$ in Eq.~(\ref{eq:stat4321}) and derive its equivalent expression (\ref{eq:stat4321_new}).  Due to Eqs.~(\ref{double_m_charge},\ref{Q_fractional_M1}), we introduce four integer numbers $k_0,k_1,k_2,k_3$ via 
   \begin{align}
   M=3k\,, \,N_m^a=2k_1\,, \,N_m^b=2k_2\,, \,Q=\frac{k_0}{6}
   \end{align} so as to simplify the analysis below. Then,  solving Eqs.~(\ref{equation_NMA},\ref{equation_NMB},\ref{Q_expression}) leads to:
\begin{align}
 &n^{f1}=-\frac{5}{2}k+\frac{k_0}{6}+\frac{5}{3}k_1-\frac{2}{3}k_2\,,\\
 &n^{f2}=-\frac{13}{2}k+\frac{k_0}{6}+\frac{11}{3}k_1+\frac{1}{3}k_2\,,\\
 &n^{f3}=-\frac{9}{2}k+\frac{k_0}{6}+\frac{8}{3}k_1-\frac{5}{3}k_2\,.
   \end{align}
Therefore, $\Gamma$ in Eq.~(\ref{eq:stat4321_new}) can be  reformulated as:
\begin{align}
\Gamma=(M+1)(-14k+\frac{1}{2}k+\frac{k_0}{2}+8k_1-2k_2)\,.
\end{align}
Since $(M+1)(-14k+8k_1-2k_2)$ is always an even integer, we may remove it and end up with:
\begin{align}
\Gamma=&(M+1)(\frac{1}{2}k+\frac{k_0}{2})\, 
=3(M+1)(Q+\frac{1}{18}M)
\end{align}
which can be rewritten as:
\begin{align}
\Gamma=&3(M+1)(Q-\frac{\Theta}{2\pi}M)
\end{align}
with $\Theta=-\frac{1}{9}\pi$. One can check that $\Gamma$ is invariant under the shift $\Theta\rightarrow \Theta+\frac{2}{9}\pi$ since the additional term $-3(M+1)\frac{1}{9}M=-k(3k+1)$ is always an even integer which leaves the even / odd property of $\Gamma$ unaltered. From this point of view, we say that two $\Theta$'s are topologically equivalent if their difference is given by multiples of $\frac{2}{9}\pi$. In conclusion, 
\begin{align}
\Theta=\frac{1}{9}\pi \text{  mod  }\frac{2}{9}\pi\,.
\end{align}
As a result, $-\frac{1}{9}\pi$, $\frac{1}{9}\pi$, $-\frac{5}{9}\pi$, etc. describe the same {FTI} states. The periodicity $\frac{2}{9}\pi$ is the minimal one in the sense that any shift smaller than $\frac{2}{9}\pi$ does not  keep the even-odd property of $\Gamma$ invariant. In other words, the charge lattice with ``tilt angle'' $\Theta=\frac{1}{9}\pi$ is always different from a lattice with $\Theta=0$. This periodicity check is very important since it is possible that a nonzero $\Theta$  might be entirely removed by a periodic shift. If this happens, the resulting bulk state is actually a trivial state. 

Next, we calculate the lattice sites in Fig.~\ref{figure_theta}(b). Since $Q=\frac{1}{6}$ and $M=3$, we have: $k_0=1,k=1$:
\begin{align}
 &n^{f1}=(-2+2k_1-k_2)+\frac{-1-k_1+k_2}{3}\,,\\
 &n^{f2}=(-6+4k_1)+\frac{-1-k_1+k_2}{3}\,,\\
 &n^{f3}=(-4+3k_1-2k_2)+\frac{-1-k_1+k_2}{3}\,.
    \end{align}
Therefore, $-1-k_1+k_2$ should be quantized in multiples of 3 such that the  $n^{fi}$'s are integer-valued. By noting that $N^a_m=2k_1,N^b_m=2k_2$, we end up with  Fig.~\ref{figure_theta}(c) where $k_0=1,k=0$ are assumed.

\section{Technical details in Sec.~\ref{sec:charlos_defect_symmetry}}\label{appendix_proof_group}

\subsection{$\mathsf{Charles}$ is a group}

\begin{proof}

\underline{\emph{Step-1}} is to prove that $\mathsf{Auto}(K)$ is a group. In other words, the elements satisfy the four group axioms. (identity element, inverse element, closure, associativity). 

Identity element.--- The identity element is $\mathscr{G}_{\mathbb{I}}=W\oplus \Omega=\mathbb{I}\oplus\mathbb{I}$ where $\mathbb{I}$ is a rank-2 identity matrix.  For every element $\mathscr{G}$ in $\mathsf{Auto}(K)$, the equation $\mathscr{G}\cdot \mathscr{G}_{\mathbb{I}}=\mathscr{G}_{\mathbb{I}} \cdot\mathscr{G}=\mathscr{G} $ holds. Here the symbol $\cdot$ denotes matrix multiplication. We will also omit it unless otherwise specified.  

Associativity.--- Associativity is guaranteed by matrix multiplication rules.

Inverse element.--- The inverse element of $\mathscr{G}$ is given by: $\mathscr{G}^{-1}=W^{-1}\oplus \Omega^{-1}$. One may check that $(W^{-1}\oplus \Omega^{-1})\cdot (W\oplus \Omega)=\mathbb{I}\oplus\mathbb{I}=\mathscr{G}_{\mathbb{I}}$ and $ (W\oplus \Omega)\cdot (W^{-1}\oplus \Omega^{-1})=\mathscr{G}_{\mathbb{I}}$, which means that: $\mathscr{G}^{-1}\cdot\mathscr{G}=\mathscr{G}\cdot\mathscr{G}^{-1}=\mathscr{G}_{\mathbb{I}}$.

Closure.--- Suppose $\mathscr{G}'=W'\oplus \Omega'\in \mathsf{Auto}(K)$. Thus, $W,\Omega, W',\Omega'$ matrices satisfy conditions (i) and (ii) in Definition~\ref{dfn_charlos}. Then, by definition, $\mathscr{G}'\cdot \mathscr{G}=(W'  W)\oplus (\Omega'  \Omega)$. Both $W'  W$ and $\Omega'  \Omega$ are still rank-2 unimodular matrices. Furthermore, 
\begin{align}(\Omega'\Omega)K (W' W)^T=  \Omega'(\Omega K W^T) {W'}^T=  \Omega'K{W'}^T=K\,.\nonumber
\end{align}
 $\therefore$ condition (i) is satisfied. And,  
 \begin{align}
 &\Gamma (\cdots,(W'W)^{-1}\mathbf{N}_m,\cdots)\nonumber\\
 =&\Gamma (\cdots,W^{-1}{W'}^{-1}\mathbf{N}_m,\cdots)\nonumber\\
 =&\Gamma (\cdots,{W'}^{-1}\mathbf{N}_m,\cdots)\nonumber\\
 =&\Gamma (\cdots,\mathbf{N}_m,\cdots)\,.\nonumber
 \end{align} $\therefore$ condition (ii) is also satisfied.

\underline{\emph{Step-2}} is to verify that $\mathsf{Inner}(K)$ is a subgroup of $\mathsf{Auto}(K)$. First, it is a subset of $\mathsf{Auto}(K)$, i.e., $\mathsf{Inner}(K)\subset\mathsf{Auto}(K)$ since not only do $W$ and $\Omega$ satisfy condition (i) and (ii), but also satisfy  
\begin{align}
W^{-1}\mathbf{N}_m-\mathbf{N}_m=K^T\,(n_1,n_2)^T
\end{align}
  and 
  \begin{align}\Omega^{-1}\mathbf{L}-\mathbf{L}=K(n_3,n_4)^T\,.
\end{align}  
    Here $n_1,\cdots,n_4$ are integers. $n_1,n_2$ depend on $\mathbf{N}_m,W$; $n_3,n_4$ depend on $\mathbf{L},\Omega$.  

      $\mathscr{G}_{\mathbb{I}}\in \mathsf{Inner}(K)$ since one can obtain $\mathscr{G}_{\mathbb{I}}\mathbf{L}-\mathbf{L}=0$ and $\mathscr{G}_{\mathbb{I}}\mathbf{N}_m-\mathbf{N}_m=0$ by choosing $n_1=n_2=n_3=n_4=0$ for all $\mathbf{L}$'s and $\mathbf{N}_m$'s.  
Then,   elements of $\mathsf{Inner}(K)$   have associativity arising from the standard matrix multiplication. 

 For the existence of the inverse element, we need to prove that $W^{-1}\oplus\Omega^{-1}\in\mathsf{Inner}(K)$.  By definition, for $W\oplus \Omega\in\mathsf{Inner}(K)$,  the operation of $W$  is \begin{align}
 W^{-1}\mathbf{N}_m-\mathbf{N}_m=K^T \mathbf{J}_{W,\mathbf{N}_m}\,,\nonumber
 \end{align} where the notation $\mathbf{J}_{W,\mathbf{N}_m}$ denotes the integer vector $(n_1,n_2)^T$ and the subscript $W,\mathbf{N}_m$ means that the integer vector is a function of $W$ and $\mathbf{N}_m$. Likewise, we also have: 
 \begin{align}
 W^{-1} (W\mathbf{N}_m)-W\mathbf{N}_m=K^T \mathbf{J}_{W,W\mathbf{N}_m}.\nonumber
 \end{align} As a result,  $W\mathbf{N}_m-\mathbf{N}_m=-K^T \mathbf{J}_{W,W\mathbf{N}_m}$. Since $-\mathbf{J}_{W,W\mathbf{N}_m}$ is an integer vector, we obtain the operation of $W^{-1}$:    
 \begin{align}
 (W^{-1})^{-1}\mathbf{N}_m-\mathbf{N}_m=K^T (-\mathbf{J}_{W,W\mathbf{N}_m})\nonumber
 \end{align} which exactly satisfies the defining property of $\mathsf{Inner}(K)$. Likewise, we can also prove that $\Omega^{-1}$ is an operation in $\mathsf{Inner}(K)$. Therefore, the inverse element $\mathscr{G}^{-1}=W^{-1}\oplus \Omega^{-1}\in\mathsf{Inner}(K)$.

For the closure property, we need to prove that $\mathscr{G}'\cdot\mathscr{G}\in \mathsf{Inner}(K)$ if   $\mathscr{G}'\in \mathsf{Inner}(K)$ and $\mathscr{G}\in \mathsf{Inner}(K)$. For this purpose, let us calculate: 
\begin{align}
&(W'W)^{-1}\mathbf{N}_m-\mathbf{N}_m \nonumber\\
=&W^{-1}{W'}^{-1}\mathbf{N}_m-\mathbf{N}_m=W^{-1}\left( \mathbf{N}_m+K^T\mathbf{J}_{W',\mathbf{N}_m}\right)\!-\!\mathbf{N}_m\nonumber\\
=&\!\left(W^{-1} \mathbf{N}_m\!-\!\mathbf{N}_m\right)\!+W^{-1}K^T\mathbf{J}_{W',\mathbf{N}_m}\nonumber\\
=&K^T\! \mathbf{J}_{W,\mathbf{N}_m}\! +\!K^T\Omega^T\mathbf{J}_{W',\mathbf{N}_m}=K^T\!\left(\mathbf{J}_{W,\mathbf{N}_m}\!+\Omega^T\mathbf{J}_{W',\mathbf{N}_m}\!\right)\,,\nonumber
\end{align}
where we have used $W^{-1}K^T=K^T\Omega^T$ that is an equivalent expression of   condition (i).
 Since $\mathbf{J}_{W,\mathbf{N}_m}+\Omega^T\mathbf{J}_{W',\mathbf{N}_m}$ is an integer vector (by noting that $\Omega$ is unimodular), we conclude that $W'W$ satisfies the defining property of $\mathsf{Inner}(K)$. So does $\Omega\Omega'$. Therefore, $\mathscr{G}'\cdot\mathscr{G}\in \mathsf{Inner}(K)$.

\underline{\emph{Step-3}} is to prove that $\mathsf{Inner}(K)$ is normal. In other words, we need to verify that $W'W{W'}^{-1}\in\mathsf{Inner}(K)$ and $\Omega'\Omega{\Omega'}^{-1}\in\mathsf{Inner}(K)$ for $\forall W,\Omega \in\mathsf{Inner}(K)$ and $\forall W',\Omega'\in\mathsf{Auto}(K)$. For this  purpose, let us calculate:
\begin{align}
&(W'W{W'}^{-1})^{-1}\mathbf{N}_m-\mathbf{N}_m=W'(W^{-1}{W'}^{-1}\mathbf{N}_m)-\mathbf{N}_m\nonumber\\
=&W'( {W'}^{-1}\mathbf{N}_m+K^T\mathbf{J}_{W,{W'}^{-1}\mathbf{N}_m})-\mathbf{N}_m\nonumber\\
=&W'K^T\mathbf{J}_{W,{W'}^{-1}\mathbf{N}_m}.\nonumber
\end{align} Since $W'\in\mathsf{Auto}(K)$, condition (i) in Eq.~(\ref{eqn:auto}) leads to: $\Omega' K{W'}^{T}=K,$ and thereby ${W'}K^T{\Omega'}^{T}=K^T$.  Therefore, ${W'}K^T=K^T({\Omega'}^{T})^{-1}$. $\therefore$
\begin{align}
\!\!\!\!(W'W{W'}^{-1})^{-1}\mathbf{N}_m-\mathbf{N}_m\!=\!K({\Omega'}^{T})^{-1} \mathbf{J}_{W,{W'}^{-1}\mathbf{N}_m}.\nonumber
\end{align} 

Since $({\Omega'}^{T})^{-1}$ is obviously a unimodular matrix, it implies that $({\Omega'}^{T})^{-1} \mathbf{J}_{W,{W'}^{-1}\mathbf{N}_m}$ is an integer vector. Thus, by the definition of $\mathsf{Inner}(K)$, $W'W{W'}^{-1}\in\mathsf{Inner}(K)$. Likewise, we also have $\Omega'\Omega{\Omega'}^{-1}\in\mathsf{Inner}(K)$. $\therefore$ we conclude that  $\mathsf{Inner}(K)$ is a normal subgroup of $\mathsf{Auto}(K)$.

 Then, according to the definition of  $\mathsf{Charles}$, the elements of $\mathsf{Charles}$ form a quotient group of $\mathsf{Auto}(K)$ by $\mathsf{Inner}(K)$. It can be non-Abelian since $\mathscr{G}\cdot \mathscr{G}'\neq \mathscr{G}'\cdot \mathscr{G}$ may hold for some elements.
\end{proof}

 \subsection{Graphical representations of $\mathsf{Charles}$ transformations}
\begin{figure}[t]
\centering
\includegraphics[width=8.5cm]{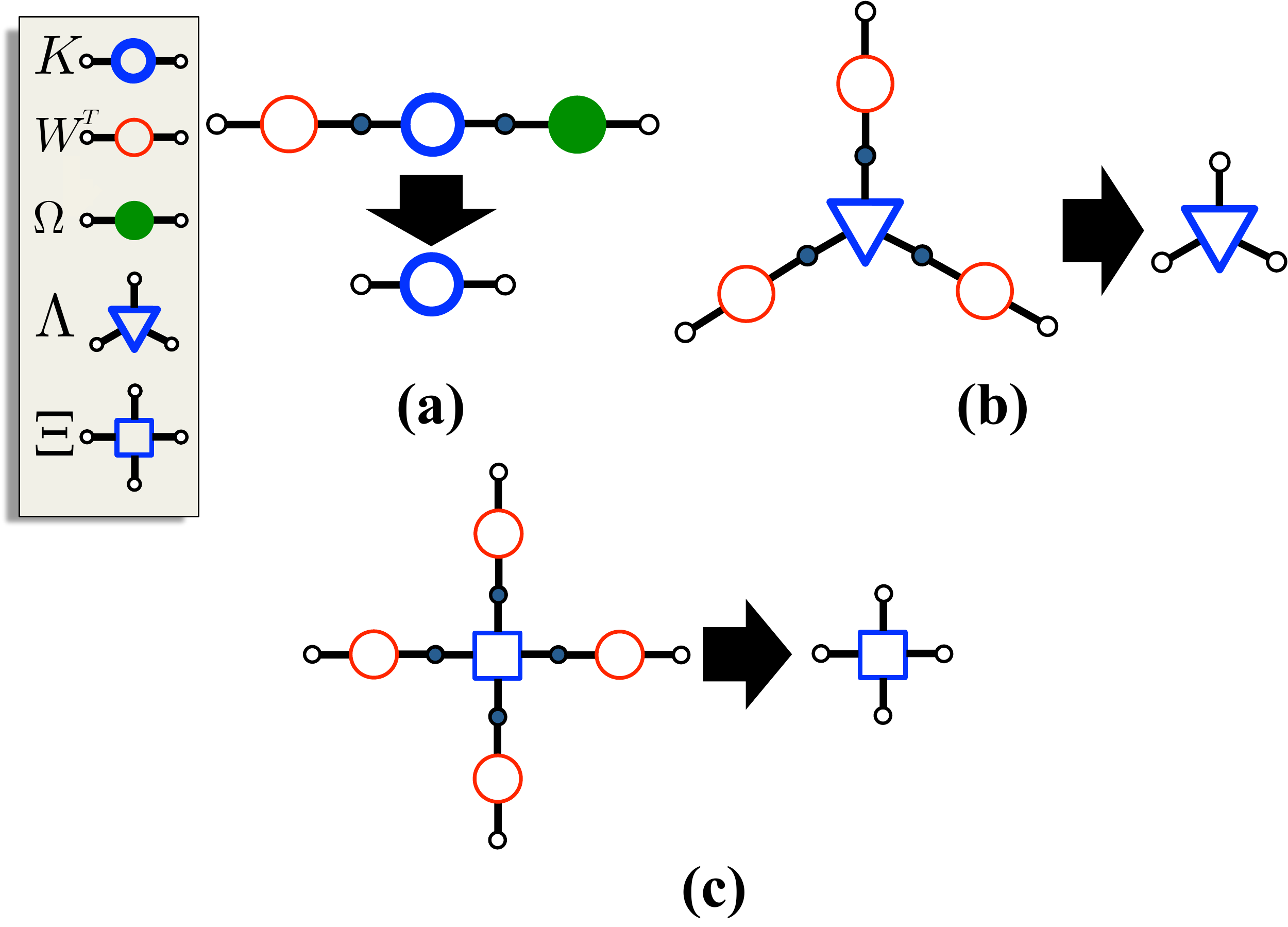}
\caption{(Color online) Tensor-network-type graphical representations of $\mathsf{Charles}$ transformations. (a) represents Eq.~(\ref{eqn:auto}) where $K$ is a fixed-point matrix. (b) represents Eq.~(\ref{eqn:auto33}) where $\Lambda$ is a fixed-point tensor with a bond dimension no less than two. (c) represents Eq.~(\ref{eqn:auto44}) where $\Xi$ is a fixed-point tensor with a bond dimension  no less than four.}
\label{figure_tensor}
\end{figure}

Eqs.~(\ref{eqn:auto},\ref{eqn:auto33},\ref{eqn:auto44}) are graphically represented in Fig.~\ref{figure_tensor} where a tensor-network-type graph is introduced. 

%

\end{document}